\documentclass[12pt]{article}
\pdfoutput=1

\usepackage{hyperref}

\usepackage{cite}
\usepackage{appendix}
\usepackage{epsfig}
\usepackage{amsmath}
\usepackage{amssymb}
\usepackage{bm}
\usepackage{hyperref}
\usepackage{slashed}
\usepackage[normalem]{ulem} 
\newcommand{\be} {\begin{equation}}
\newcommand{\ee} {\end{equation}}
\newcommand{\beq} {\begin{equation}}
\newcommand{\eeq} {\end{equation}}
\newcommand{\ba} {\begin{eqnarray}}
\newcommand{\ea} {\end{eqnarray}}
\newcommand{\vareps}{\varepsilon}

\newcommand{\Lag} {\mathcal L}
\newcommand{\cA} {\mathcal A}
\newcommand{\cT} {\mathcal T}
\newcommand{\cd} {{\cdot }}

\newcommand{\GeV}{\text{GeV}}
\newcommand{\TeV}{\text{TeV}}

\usepackage{xspace}
\newcommand{\ord}{\mathcal{O}}

\newcommand{\OpenLoops}{{\rmfamily\scshape OpenLoops}\xspace}
\newcommand{\Sherpa}{{\rmfamily\scshape Sherpa}\xspace}
\newcommand{\SherpaOpenLoops}{{\rmfamily\scshape Sherpa+OpenLoops}\xspace}
\newcommand{\MadGraph}{{\rmfamily\scshape MadGraph5\_aMC@NLO}\xspace}
\newcommand{\FeynRules}{{\rmfamily\scshape FeynRules}\xspace}
\newcommand{\UFO}{{\rmfamily\scshape UFO}\xspace}

\newcommand{\kT}{k_{\mathrm{T}}}
\newcommand{\pT}{\ensuremath{p_\mathrm{T}}\xspace}
\newcommand{\pTj}{\ensuremath{p_\mathrm{T,j}}\xspace}
\newcommand{\pTjone}{\ensuremath{p_\mathrm{T,j_1}}\xspace}
\newcommand{\pTjtwo}{\ensuremath{p_\mathrm{T,j_2}}\xspace}
\newcommand{\HThalf}{\ensuremath{H_\mathrm{T}/2}\xspace}
\newcommand{\pTZ}{\ensuremath{p_\mathrm{T,Z}}\xspace}
\newcommand{\pTW}{\ensuremath{p_\mathrm{T,W}}\xspace}
\newcommand{\pTV}{\ensuremath{p_\mathrm{T,V}}\xspace}
\newcommand{\pTH}{\ensuremath{p_\mathrm{T,H}}\xspace}
\newcommand{\minvZH}{\ensuremath{m_\mathrm{HZ}}\xspace}
\newcommand{\rF}{\mathrm{F}}
\newcommand{\rR}{\mathrm{R}}

\newcommand{\VH}{VH}
\newcommand{\Vh}{$Vh$\ }

\usepackage{color}
\definecolor{darkblue}{cmyk}{1,0.3,0,0.2}
\definecolor{violet}{cmyk}{0,1,0,0.2}
\hypersetup{colorlinks, bookmarksnumbered, citecolor=darkblue, linkcolor=darkblue, pdfstartview=FitH, urlcolor=darkblue, linktocpage}

\begin{document}

\begin{flushright}
 ZU-TH-47/15 \\
 December 2015
\end{flushright}

\thispagestyle{empty}

\bigskip

\begin{center}
\vspace{1.5cm}
    {\Large\bf  Pseudo-observables in electroweak Higgs production} \\[1cm]
   {\bf Admir Greljo$^{a,c}$, Gino Isidori$^{a,b}$, \\[0.1cm]
   Jonas M. Lindert$^a$, David Marzocca$^a$}    \\[0.5cm]
  {\em $(a)$  Physik-Institut, Universit\"at Z\"urich, CH-8057 Z\"urich, Switzerland}  \\ 
   {\em $(b)$  INFN, Laboratori Nazionali di Frascati, I-00044 Frascati, Italy}\\
  {\em $(c)$  Faculty of Science, University of Sarajevo, Zmaja od Bosne 33-35, 71000 Sarajevo, Bosnia and Herzegovina }  \\[1.0cm]
\end{center}

\centerline{\large\bf Abstract}
\begin{quote}
\indent
We discuss how the leading electroweak Higgs production processes at the LHC, 
namely vector-boson fusion and Higgs+$W/Z$ associated production,
can be characterized in generic extensions of the Standard Model by a proper set of pseudo-observables (PO).
We analyze the symmetry properties of these PO and their relation with the PO set  appearing in Higgs decays.
We discuss in detail the kinematical studies necessary to extract the production PO from data, and present a
first estimate of the LHC sensitivity on these observables in the high-luminosity phase. 
The impact of QCD corrections and the kinematical studies necessary to test 
the validity of the momentum expansion at the basis of the PO decomposition are also discussed. 
\end{quote}
\vspace{5mm}

\newpage

\tableofcontents

\section{Introduction}

Characterizing the properties of the Higgs boson, both in production and in
decay processes, with high precision and minimum theoretical bias, is one of the
main goal of future experimental efforts in high-energy physics and a promising
avenue to shed light on physics beyond the Standard Model (SM). In this context,
a useful tool is provided by the so-called Higgs pseudo-observables
(PO)~\cite{Gonzalez-Alonso:2014eva,Gonzalez-Alonso:2015bha,Ghezzi:2015vva,Passarino:2010qk,David:2015waa}.
The latter constitute a finite set of parameters that are experimentally accessible, are
well-defined from the point of view of quantum field theory (QFT), and
characterize possible deviations from the SM in processes involving the Higgs
boson in great generality. More precisely, the Higgs PO are defined from a
general decomposition of on-shell amplitudes involving the Higgs boson --based
on analyticity, unitarity, and crossing symmetry-- and a momentum expansion
following from the dynamical assumption of no new light particles (hence no
unknown physical poles in the amplitudes) in the kinematical regime where the
decomposition is assumed to be valid.

The idea of PO has been formalized the first time in the context of electroweak
observables around the $Z$ pole~\cite{Bardin:1999gt,ALEPH:2005ab}, while the
generalization relevant to analyze Higgs decays has been presented in
Ref.~\cite{Gonzalez-Alonso:2014eva}. In this paper we further generalize the PO
approach to describe electroweak Higgs-production processes, namely
vector-boson fusion (VBF) and associated production with a massive SM gauge boson
(\VH).

The interest of such production processes is twofold. On the one hand, they are
closely connected to the $h\to4\ell,2\ell2\nu$ decay processes by crossing
symmetry, and by the exchange of lepton currents into quark currents. As a
result, some of the Higgs PO necessary to describe the $h\to4\ell,2\ell2\nu$
decay kinematics appear also in the description of the VBF and \VH~cross
sections (independently of the Higgs decay mode). This fact opens the
possibility of combined analyses of production cross sections and differential
decay distributions, with a significant reduction on the experimental error on
the extraction of the PO. On the other hand, studying the production cross sections allows
us to explore different kinematical regimes compared to the decays. By
construction, the momentum transfer appearing in the Higgs decay amplitudes is
limited by the Higgs mass, while such limitation is not present in the
production amplitudes. This fact allows us to test the momentum expansion that is
intrinsic in the PO decomposition, as well as in any effective field theory
approach to physics beyond the SM.

Despite the similarities at the fundamental level, the phenomenological
description of VBF and \VH~in terms of PO is significantly more challenging
compared to that of Higgs decays. On the one hand, QCD corrections play a
non-negligible role in the production processes. Although technically
challenging, this fact does not represent a conceptual problem for the PO
approach: the leading QCD corrections factorize in VBF and \VH, similarly to the
factorization of QED corrections in $h\to 4\ell$~\cite{Bordone:2015nqa}. As we
will show, this implies that NLO QCD corrections can be incorporated in general
terms with suitable modifications of the existing Monte Carlo tools. On the
other hand, the relation between the kinematical variables at the basis of the
PO decomposition (i.e.~the momentum transfer of the partonic currents, $q^2$)
and the kinematical variables accessible in $pp$ collisions is not
straightforward, especially in the VBF case. As we will show, this problem finds
a natural solution in the VBF case due to strong correlation between $q^2$ and
the $\pT$ of the VBF-tagged jets.

The paper is organized as follows: in Section~\ref{sect:ampdec} we present the
decomposition in terms of PO of the electroweak amplitudes relevant to VBF and
\VH, analyzing the relation with the decay PO already introduced in
Ref.~\cite{Gonzalez-Alonso:2014eva}. In Section~\ref{sect:VBF} we present a
phenomenological analysis of the VBF process, discussing in detail the
implementation of QCD corrections, and the key role of the jet $\pT$ for the
identification of the PO. An estimate of the statistical error expected on the
PO extracted from VBF in the high-luminosity phase at the LHC is also presented.
A similar discussion for the \VH~processes is presented in
Section~\ref{sect:VH}. A detailed discussion about the validity of the momentum
expansion, and how to test it from data, is presented in
Section~\ref{sect:expansion}. The results are summarized in the Conclusions.

\section{Amplitude decomposition}
\label{sect:ampdec}

Neglecting  light fermion masses, the electroweak production processes \VH~and VBF or, 
more precisely, the electroweak partonic amplitudes $f_1  f_2  \to h +  f_3  f_4$, 
can be completely described by the three-point correlation function of the Higgs boson and two (color-less) fermion currents
\be
	\langle 0 | \cT\left\{ J_f^\mu(x), J_{f^\prime}^\nu (y), h(0) \right\}| 0\rangle ~,
	\label{eq:corr_func}
\ee
where all the states involved are on-shell. The same correlation function also controls also four-fermion Higgs decays. 
In the  $h\to4\ell,2\ell2\nu$ case both currents are leptonic and all fermions are in the final state~\cite{Gonzalez-Alonso:2014eva}. In case of \VH~associate production one of the currents describes the initial state quarks, while the other describes the decay products of the (nearly on-shell) vector boson. Finally, in VBF production the currents are not in the $s$-channel as in the previous cases, but in the $t$-channel. 
Strictly speaking, in \VH~and VBF  the quark states  are not on-shell; however, 
their off-shellness of order $\Lambda_{\text{QCD}}$ can be safely neglected compared to the 
electroweak scale   characterizing the hard process (both within and beyond the SM).

Following Ref.~\cite{Gonzalez-Alonso:2014eva}, we expand the correlation
function in Eq.~\eqref{eq:corr_func} around the known physical poles due to the
propagation of intermediate SM electroweak gauge bosons. The PO are then defined
by the residues on the poles and by the non-resonant terms in this expansion. By
construction, terms corresponding to a double pole structure are independent
from the nature of the fermion current involved. As a result, the corresponding
PO are universal and can be extracted from any of the processes mentioned
above, both in production and in decays.

\subsection{Vector boson fusion Higgs production}
\label{sec:VBFdec}

Higgs production via vector boson fusion (VBF) receives contribution both from neutral- and charged-current channels. Also, depending on the specific partonic process, there might be two different ways to construct the two currents, and these two terms interfere with each other. For example, in $u u \to u u h$ two neutral-current processes interfere, while in $u d \to u d h$ there is an interference between neutral and charged currents. In this case it is clear that one should sum the two amplitudes with the proper symmetrization, as done in the case of $h\to4e$~\cite{Gonzalez-Alonso:2014eva}. 

We now proceed describing how each of these amplitudes can be parametrized in terms of PO.
Let us start with the neutral-current one.
The amplitude for the on-shell process $q_i(p_1) q_j(p_2) \to q_i(p_3) q_j(p_4) h(k)$ can be parametrized by
\beq
	\cA_{n.c}\left(q_i(p_1) q_j(p_2) \to q_i(p_3) q_j(p_4) h(k) \right) = i \frac{2m_Z^2}{v} \bar{q_i}(p_3) \gamma_\mu q_i(p_1) \bar{q_j}(p_4) \gamma_\nu q_j(p_2) \cT^{\mu\nu}_{n.c.}(q_1, q_2) ,
	\label{eq:VBFnc}
\eeq
where $q_1 = p_1 - p_3$, $q_2 = p_2 - p_4$ and $\cT^{\mu\nu}_{n.c.}(q_1, q_2)$ is the same tensor structure appearing in $h \to 4f$ decays~\cite{Gonzalez-Alonso:2014eva}. In particular, Lorentz invariance allows only three possible tensor structures, to each of which we can assign  a generic form factor:
\be
\cT^{\mu\nu}_{n.c.} (q_1, q_2) =  \left[ F^{q_i q_j}_L (q_1^2, q_2^2) g^{\mu\nu} +  F^{q_i q_j}_T (q_1^2, q_2^2)  \frac{ {q_1}\cd {q_2}~g^{\mu\nu} -{q_2}^\mu {q_1}^\nu }{m_Z^2}  
 +   F^{q_i q_j}_{CP} (q_1^2, q_2^2)  \frac{  \vareps^{\mu\nu\rho\sigma} q_{2\rho} q_{1\sigma}   }{m_Z^2}  \right]~.
\ee
The form factor $F_L$ describes the interaction with the longitudinal part of
the current, as in the SM; the $F_T$ term describes the interaction with the
transverse part, while $F_{CP}$ describes the CP-violating part of the
interaction (if the Higgs is assumed to be a CP-even state).

The charged-current contribution to the amplitude for the on-shell process $u_i(p_1) d_j(p_2) \to d_k(p_3) u_l(p_4) h(k)$ can be parametrized by
\beq
	\cA_{c.c}\left(u_i(p_1) d_j(p_2) \to d_k(p_3) u_l(p_4) h(k) \right) = i \frac{2m_W^2}{v} \bar{d_k}(p_3) \gamma_\mu u_i(p_1) \bar{u_l}(p_4) \gamma_\nu d_j(p_2) \cT^{\mu\nu}_{c.c.}(q_1, q_2) ,
	\label{eq:VBFcc}
\eeq
where, again, $\cT^{\mu\nu}_{c.c.}(q_1, q_2)$ is the same tensor structure 
appearing in the charged-current $h \to 4f$ decays:
\be
	\cT_{c.c.}^{\mu\nu} (q_1, q_2) =  \left[ G^{ijkl}_L (q_1^2, q_2^2) g^{\mu\nu} +  G^{ijkl}_T (q_1^2, q_2^2)  \frac{ {q_1}\cd {q_2}~g^{\mu\nu} -{q_2}^\mu {q_1}^\nu }{m^2_W}  
 +   G^{ijkl}_{CP} (q_1^2, q_2^2)  \frac{  \vareps^{\mu\nu\rho\sigma} q_{2\rho} q_{1\sigma}   }{m^2_W}  \right]
\ee
The amplitudes for the processes with  anti-quarks in the initial state can easily be obtained from the above ones.

The next step  in the decomposition of the amplitude requires to perform a
momentum expansion of the form factors around the physical poles due to the
propagation of SM electroweak gauge bosons ($\gamma$, $Z$ and $W^{\pm}$), and to
define the PO (i.e.~the set $\{\kappa_i, \epsilon_i\}$) from the residues of
such poles. We stop this expansion neglecting terms which can be generated only
by local operators with dimension higher than six. A discussion about
limitations and consistency checks of this procedure is presented in
Section~\ref{sect:expansion}.
The explicit form of the expansion of all the form factors in term of PO 
can be found in Ref.~\cite{Gonzalez-Alonso:2014eva}\footnote{~With respect to \cite{Gonzalez-Alonso:2014eva} we modified the labels of the form factors: $F_1 \to F_L$, $F_3 \to F_T$ and $F_4 \to F_{CP}$, and analogously for the $G_i$.}
and will not be repeated here. 
We report here explicitly only expression  for the longitudinal form factors,
 which are the only ones containing PO not present also in the decay amplitudes:
\be
\begin{split}
F^{q_i q_j}_L (q_1^2, q_2^2) &= \kappa_{ZZ}  \frac{ g_Z^{q_i}  g_Z^{q_j}  }{P_Z(q_1^2) P_Z(q_2^2)}
  +  \frac{\epsilon_{Z q_i}}{m_Z^2}  \frac{ g_Z^{q_j}   }{  P_Z(q_2^2)} + \frac{\epsilon_{Z q_j}}{m_Z^2}   \frac{ g_Z^{q_i}   }{  P_Z(q_1^2)}  + \Delta^{\rm SM}_{L, n.c.} (q_1^2, q_2^2) ~, \\
G^{ijkl}_L (q_1^2, q_2^2) &= \kappa_{WW} \frac{ g_W^{ik}  g_W^{jl}  }{P_W(q_1^2) P_W(q_2^2)}
  +  \frac{\epsilon_{W ik}}{m_W^2}  \frac{ g_W^{jl}   }{  P_W(q_2^2)} + \frac{\epsilon_{W jl}}{m_W^2}   \frac{ g_W^{ik}   }{  P_W(q_1^2)}  + \Delta^{\rm SM}_{L, c.c} (q_1^2, q_2^2) ~.
\end{split}
\label{eq:FLGL}
\ee
Here $P_V(q^2) = q^2 - m_V^2 + i m_V \Gamma_V$, while $g_Z^f$ and $g^{ik}_{W}$
are the PO characterizing the on-shell couplings of $Z$ and $W$ boson to a pair
of fermions: within the SM $g_Z^f = \frac{g}{c_{\theta_W}} (T_3^f - Q_f
s_{\theta_W}^2)$ and $g^{ik}_{W} = \frac{g}{\sqrt{2}} V_{ik}$, where $V$ is the
CKM mixing matrix and $s_{\theta_W}$ ($c_{\theta_W}$) is the sine (cosine) of the Weinberg angle.\footnote{~More precisely, $(g^{ik}_{W})_{\rm SM}
=\frac{g}{\sqrt{2}} V_{ik}$ if $i$ and $k$ refers to left-handed quarks,
otherwise $(g^{ik}_{W})_{\rm SM}=0$.} The functions $\Delta^{\rm SM}_{L, n.c.
(c.c.)} (q_1^2, q_2^2)$ denote non-local contributions generated at the one-loop
level (and encoding multi-particle cuts) that cannot be re-absorbed into the
definition of $\kappa_i$ and $\epsilon_i$. At the level of precision we are
working, taking into account also the high-luminosity phase of the LHC, these
contributions can be safely be fixed to their SM values.

As anticipated, the crossing symmetry between $h\to 4 f$ and $2 f \to h \, 2 f$
amplitudes ensures that the PO are the same in production and decay (if the same
fermions species are involved). The amplitudes are explored in different
kinematical regimes in the two type of processes (in particular the
momentum-transfer, $q_{1,2}^2$, are space-like in VBF and time-like in $h\to
4f$). However, this does not affect the definition of the PO. This implies that
the fermion-independent PO associated to a double pole structure, such as
$\kappa_{ZZ}$ and $\kappa_{WW}$ in Eq.~(\ref{eq:FLGL}), are expected to be
measured with higher accuracy in $h\to 4 \ell$ and $h\to 2\ell 2\nu$ rather than
in VBF. On the contrary, VBF is particularly useful to constrain the
fermion-dependent contact terms $\epsilon_{Z q_i}$ and $\epsilon_{Wu_i d_j}$,
that appear only in the longitudinal form factors. For this reason, in the
following phenomenological analysis we focus our attention mainly on the LHC
reach on these parameters.  Still, we  stress  that the PO framework is well suited
to perform a global fit including production and decay observables at the same time. 

\subsection{Associated vector boson plus Higgs production}
\label{sec:VHprod_ff}

By \VH~we denote the production of the Higgs boson with a nearly 
on-shell massive vector boson ($W$ or $Z$), starting from and initial $q\bar q$ state. 
For simplicity, in the following we will assume 
that the vector boson is on-shell and that the interference with
the VBF amplitude can be neglected. However, we stress that the PO formalism clearly allows to describe 
both these effects (off-shell $V$ and interference with VBF in case of $V\to \bar q q$ decay) simply 
applying the general decomposition of  neutral-  and charged-current 
amplitudes as outlined above. 

Similarly to VBF, Lorentz invariance allows us to decompose the amplitudes for the on-shell processes $q_i(p_1) \bar{q}_i(p_2) \to h(p) Z(k)$ and $u_i(p_1) \bar{d}_j(p_2) \to h(p) W^+(k)$ in three possible tensor structures: a longitudinal one, a transverse one, and a CP-odd one,
\beq\begin{split}
	\cA &\left(q_i(p_1) \bar{q}_i(p_2) \to h(p) Z(k)\right) = i \frac{2m_Z^2}{v} \bar{q_i}(p_2) \gamma_\nu q_i(p_1) \epsilon_\mu^{Z*}(k) \times \\
	&\quad \times \left[ F_{L}^{q_i Z}(q^2) g^{\mu\nu} + F_{T}^{q_i Z}(q^2) \frac{- (q\cdot k) g^{\mu\nu} + q^\mu k^\nu}{m_Z^2} + F_{CP}^{q_i Z}(q^2) \frac{ \epsilon^{\mu\nu\alpha\beta} q_\alpha k_\beta}{m_Z^2} \right]~,
	\label{eq:AmplqqZh}
\end{split}\eeq
\beq\begin{split}
	\cA & \left(u_i(p_1) \bar{d}_j(p_2) \to h(p) W^+(k)\right) = i \frac{2m_W^2}{v} \bar{d_j}(p_2) \gamma_\nu u_i(p_1) \epsilon_\mu^{W*}(k) \times \\
	&\quad \times \left[ G_{L}^{q_{ij} W}(q^2) g^{\mu\nu} +  G_{T}^{q_{ij} W}(q^2) \frac{- (q\cdot k) g^{\mu\nu} + q^\mu k^\nu}{m_W^2} + G_{CP}^{q_{ij} W}(q^2) \frac{ \epsilon^{\mu\nu\alpha\beta} q_\alpha k_\beta}{m_W^2} \right]~,
	\label{eq:AmplqqWh}
\end{split}\eeq
where $q = p_1 + p_2 = k + p$.  In the limit where we neglect the off-shellness of the final-state $V$, 
 the form factors can only depend on $q^2$. Already from this decomposition of the amplitude it  should be clear that differential measurements of the VH cross sections as a function of $q^2$~\cite{Isidori:2013cga}, as well as  in terms of  angular variables that allow to disentangle  different tensor structures, are an important input to constrain the PO.

Performing the momentum expansion of the form factors around the physical poles, and defining 
the PO as in Higgs decays and VBF, we find
\be\begin{array}{rcl rcl}
	F_{L}^{q_i Z}(q^2)	&=& \kappa_{ZZ} \frac{g_{Zq_i}}{P_Z(q^2)} + \frac{\epsilon_{Zq_i}}{m_Z^2}\,,
\qquad&	G_{L}^{q_{ij} W}(q^2)	&=& \kappa_{WW} \frac{(g^{u_i d_j}_{W})^*}{P_W(q^2)} + \frac{\epsilon_{Wu_i d_j}^*}{m_W^2}\,, \\
	F_{T}^{q_i Z}(q^2)	&=& \epsilon_{ZZ} \frac{g_{Zq_i}}{P_Z(q^2)} + \epsilon_{Z\gamma} \frac{e Q_q}{q^2}\,,
\qquad&	G_{T}^{q_{ij} W}(q^2)	&=& \epsilon_{WW} \frac{(g^{u_i d_j}_{W})^*}{P_W(q^2)}\,, \\
	F_{CP}^{q_i Z}(q^2)	&=& \epsilon^{\rm CP}_{ZZ} \frac{g_{Zq_i}}{P_Z(q^2)} - \epsilon^{\rm CP}_{Z\gamma} \frac{e Q_q}{q^2} \,,
\qquad&	G_{CP}^{q_{ij} W}(q^2) &=& \epsilon^{\rm CP}_{WW} \frac{(g^{u_i d_j}_{W})^*}{P_W(q^2)}\,,
\end{array}
	\label{eq:FFVh}
\ee
where we have omitted the indication of the (tiny) non-local terms, fixed to their corresponding SM values.   
According to the  arguments already discussed at the end of Sect.~\ref{sec:VBFdec}, in the following
phenomenological analysis we focus our attention on the longitudinal form factors $F_L$ and $G_L$
and, in particular, on the extraction of the  quark contact terms
$\epsilon_{Zq_i}$ and $\epsilon_{Wu_i d_j}$.

\subsection[Parameter counting, symmetry limits, and dynamical assumptions]{Parameter counting, symmetry limits, and dynamical assumptions on the PO}

We  now want to analyze the number of free parameters  and the symmetry limits for the newly introduced PO appearing in VBF and \VH~production, 
compared to the decay PO introduced in Ref.~\cite{Gonzalez-Alonso:2014eva}. 
The  additional set of PO  (the ``production PO") is represented by the 
contact terms for the light quarks. In a four-flavor scheme, in absence of any symmetry assumption, 
 the number of independent parameters for the neutral-current contact terms is 16 
 ($\epsilon_{Z q^{i j}}$, where $q=u_L, u_R, d_L, d_R$, and $i,j=1,2$): 
  8 real parameters for flavor diagonal terms 
  and 4 complex flavor-violating parameters. Similarly, there are 16 independent parameters in charged currents, namely the 
  8 complex  terms $\epsilon_{W u^{i}_{L} d^{j}_{L}}$ and  $\epsilon_{W u^{i}_{R} d^{j}_{R}}$. 
 
The number of independent PO   can be significantly  reduced  neglecting terms that violate the 
 $U(1)_f$ flavor symmetry acting on each of the light fermion species, $u_R$, $d_R$, $s_R$, $c_R$, $q_L^{(d)}$, and $q_L^{(s)}$,
 where $q_L^{(d,s)}$ denotes the two quark doublets in the basis where down quarks are diagonal. This symmetry 
is an exact symmetry of the SM in the limit where we neglect light quark masses. Enforcing it at the PO level is equivalent to neglecting 
terms that do not interfere with SM amplitudes in the limit of vanishing light quark masses. 
Under this (rather conservative) assumption,
the number of independent neutral-current contact terms reduces to 8 real parameters,\footnote{~Strictly speaking, having 
defined the quark doublets in the basis where  down quarks are diagonal, the $\epsilon_{Z u_L^{i j}}$ have a non vanishing 
off-diagonal component~\cite{Gonzalez-Alonso:2014eva}. However, this can be neglected for all practical purposes.}
\be
\epsilon_{Z u_R},~ \epsilon_{Z c_R},~ \epsilon_{Z d_R},~ \epsilon_{Z s_R},~ \epsilon_{Z d_L},~ \epsilon_{Z s_L},~
~ \epsilon_{Z u_L},~ \epsilon_{Z c_L},
\ee
and only 2 complex parameters in the charged-current case:
\be
\epsilon_{W u^{i}_{L} d^{j}_{L}}\equiv V_{i j} \epsilon_{W u^{j}_{L}},~\qquad \epsilon_{W u^{i}_{R} d^{j}_{R}}=0~.
\ee

A further interesting reduction of the number of parameters occurs under the assumption of an $U(2)^3$ symmetry acting on the first two generations, namely 
the maximal flavor symmetry compatible with the SM gauge group~\cite{Chivukula:1987py,DAmbrosio:2002ex,Barbieri:2011ci}. 
The independent parameters in this case reduces to six:
\be
\epsilon_{Z u_L},~\epsilon_{Z u_R},~ \epsilon_{Z d_L},~ \epsilon_{Z d_R},~ \epsilon_{W u_L}~,
\label{eq:MFV_contterms}
\ee
where $\epsilon_{W u_L}$ is complex, or five if we further neglect CP-violating contributions (in such case $\epsilon_{W u_L}$ is real).
We employ this set of assumptions ($U(2)^3$ flavor symmetry and CP conservation) in the phenomenological analysis of VBF 
and \VH~processes discussed in the rest of the paper.  Finally, we can enforce custodial symmetry 
that, as shown in \cite{Gonzalez-Alonso:2014eva}, implies
\be
\epsilon_{W u_L} = \frac{c_W}{\sqrt{2} } (\epsilon_{Z u_L}- \epsilon_{Z d_L})~,
\label{eq:custod}
\ee
reducing the number of independent PO to four in the  $U(2)^3$ case (independently of any assumption about  CP).

\begin{table}[t]
\begin{center}
\begin{tabular}{|c||c|c|c|} \hline\hline
\rule{0pt}{1.2em}%
Amplitudes/processes &  $U(2)^3$ flavor symm. & flavor non universality & CPV    \\
 \hline\hline
  \raisebox{-2pt}[0pt][7pt]{ neutral currents} & $\epsilon_{Z u_L}, \epsilon_{Z u_R}$   &   $\epsilon_{Z c_L}, \epsilon_{Z c_R}$    &    
   \\
   \raisebox{0pt}[0pt][7pt]{(VBF$_{n.c.}$+$Zh$)}  & $\epsilon_{Z d_L}, \epsilon_{Z d_R}$ &   $\epsilon_{Z s_L}, \epsilon_{Z s_R}$    &   
   \\ \hline
   \raisebox{-2pt}[0pt][7pt]{charged currents}    &  
   \raisebox{-11pt}[0pt][0pt]{ Re($\epsilon_{Wu_L}$) } & \raisebox{-4pt}[0pt][7pt]{ Re($\epsilon_{Wc_L}$)  } &  
   \raisebox{-4pt}[0pt][7pt]{ Im($\epsilon_{W u_L}$)}
   \\  
   (VBF$_{c.c.}$+$Wh$)   &      &   \multicolumn{2}{ c|}{  \raisebox{-2pt}[4pt][7pt]{ $\qquad$  Im($\epsilon_{W c_L}$)  }}  
   \\ \hline\hline
\raisebox{-2pt}[0pt][7pt]{ VBF and VH } &   $\epsilon_{Z u_L}, \epsilon_{Z u_R}$  &   $\epsilon_{Z c_L}, \epsilon_{Z c_R}$    &   
 \\ 
 \raisebox{0pt}[0pt][7pt]{  [with custodial symm.] } &  $\epsilon_{Z d_L}, \epsilon_{Z d_R}$    &    $\epsilon_{Z s_L}, \epsilon_{Z s_R}$   &  \\
 \hline\hline
 \end{tabular}
\caption{\label{tab:POsumm} Summary of the ``production PO",  namely the PO 
appearing in VBF and VH in addition to those already present in Higgs decays
(classified in Ref.~\cite{Gonzalez-Alonso:2014eva}).    In the second 
column we show the independent PO needed for a given set of amplitudes,  
assuming both CP invariance and $U(2)^3$ flavor symmetry. The additional variables needed if we relax these symmetry hypotheses
are reported in the third and fourth columns. In the 
bottom row we show the independent PO needed for a combined description of VBF and VH
under the hypothesis of custodial symmetry. The number of independent PO range from 12 (sum of the first two lines)
to 4 (bottom row, second column).}
\end{center}
\end{table}

\bigskip

As far as dynamical hypotheses are concerned, numerical constraints on the Higgs PO can be derived under the hypothesis that the Higgs particle is the massive excitation of a pure $SU(2)_L$ doublet, i.e.~within the so-called linear EFT (or SMEFT). In this framework the Higgs PO receive contributions 
from effective operators written in terms of the doublet field $H$, that contribute also to non-Higgs observables. As a result, it is possible to 
derive relations between the Higgs PO and electroweak precision observables. Limiting the attention to the (presumably dominant) 
tree-level contributions, generated by dimension-6 operators, 
the following relations can be derived~\cite{Gonzalez-Alonso:2015bha}
\ba
	\epsilon_{Z f} &=& \frac{2m_Z}{v} \left( \delta g^{Zf} - (c_\theta^2 T^3_f + s_\theta^2 Y_f) {\bf 1}_3 \delta g_{1,z} + t_\theta^2 Y_f {\bf 1}_3 \delta \kappa_\gamma \right)~,
	\nonumber\\
	\epsilon_{W f} &=& \frac{\sqrt{2} m_W}{v} \left( \delta g^{Wf} - c_\theta^2 {\bf 1}_3 \delta g_{1,z}  \right)~,
	\label{eq:cont_terms_EFT}
\ea
where $\delta g^{Zf}$ and $\delta g^{Wf}$ are the effective $Z$- and $W$- couplings to SM fermions, $\delta g_{1,z}$  and  $\delta \kappa_\gamma$ are the anomalous triple gauge couplings (aTGC),  and  $T^3_f$ and $Y_f$ are the isospin and hypercharge quantum numbers of the fermion $f$. 
Moreover, the custodial-symmetry relation~(\ref{eq:custod}) is automatically enforced at the dimension-6 level.

Recent analyses of $Z$- and $W$-pole observables within the SMEFT, with a generic flavor structure,  can be found in Ref.~\cite{Efrati:2015eaa,Berthier:2015gja}.  A combined fit to LEP-II WW and LHC Higgs signal strengths data, which removes all the flat directions in the determination of aTGC within the SMEFT has been presented in Ref.~\cite{Falkowski:2015jaa}. Combining some of these recent fits (in particular $Z$- and $W$-pole couplings
from Ref.~\cite{Efrati:2015eaa} and aTGC from Ref.~\cite{Falkowski:2015jaa}) we find the following  numerical 
constraints on the quark contact terms (within the SMEFT):
\be
\left(\begin{array}{c}
\epsilon_{Zu_{L}}\\
\epsilon_{Zu_{R}}\\
\epsilon_{Zd_{L}}\\
\epsilon_{Zd_{R}}
\end{array}\right)=\left(\begin{array}{c}
-0.010\pm0.008\\
0.012\pm0.011\\
0.023\pm0.023\\
0.018\pm0.037
\end{array}\right)~,
\ee 
where, for simplicity, we have further imposed the $U(2)^3$ flavor symmetry hypothesis. The corresponding correlation matrix  
turns out to be close to the identity matrix. 
The precise values of these results is not relevant to the present analysis, but it can be used as a guideline for the sensitivity needed on the 
PO measured from VBF and \VH~  in order to tests SMEFT predictions. As we show later on, the  LHC  at high luminosity will reach 
such sensitivity. 

A further restrictive dynamical hypothesis  is obtained within the framework of the so-called universal theories, i.e.~by assuming that all new physics interactions can be written in terms of the SM bosonic fields only. All analyses of VBF and \VH~production, as well as of $h\to 4f$ decays, 
performed assuming new physics only via modified $h VV$ vertices belong to this category, e.g. in Refs.~\cite{Artoisenet:2013puc,Maltoni:2013sma,Wells:2015uba,Mimasu:2015nqa}.
A specific example of this scenario are the parametric expressions of the Higgs PO in terms of
the so-called ``Higgs characterization framework" introduced in Refs.~\cite{Artoisenet:2013puc,Maltoni:2013sma}:\footnote{~We note that there is a typo for the $ \kappa_{H\partial \gamma}$ operator in \cite{Maltoni:2013sma}, while it is reported correctly in Ref.~\cite{Artoisenet:2013puc}.}
\be\begin{split}
	\kappa_{ZZ} &= c_\alpha \kappa_{\rm SM} + \frac{v c_\alpha}{\Lambda} \kappa_{H\partial Z} ~, \\
	\kappa_{WW} &= c_\alpha \kappa_{\rm SM} + \frac{v c_\alpha}{\Lambda} \kappa_{H\partial W}  ~, \\
	\epsilon_{Z f} &= \frac{g}{c_W} \left( T_f^3 - Q_f s_W^2 \right) \frac{v c_\alpha}{2 \Lambda} \kappa_{H\partial Z} + e Q_f \frac{v c_\alpha}{2 \Lambda} \kappa_{H\partial \gamma} ~, \\
	\epsilon_{W f} &= \frac{g}{\sqrt{2}} \frac{v c_\alpha}{2 \Lambda} \kappa_{H\partial W}  ~.
	\label{eq:HPO_HiggsChar}
\end{split}
\ee
 In this case the variability of the neutral-current contact terms is further reduced by a dynamical assumption that links them to the  
 two terms $\kappa_{H\partial Z}$ and $\kappa_{H\partial \gamma}$.
We stress that 
 such an assumption cannot be justified only in terms of symmetry principles.

Using \FeynRules~\cite{Alloul:2013bka} we implemented a general \UFO model \cite{Degrande:2011ua} containing all the Higgs PO (including also decays~\cite{Gonzalez-Alonso:2014eva}). The model itself will promptly be made available online~\cite{tool-link} and allows for comprehensive phenomenological Monte Carlo studies at the LHC. A detailed implementation of the Higgs PO framework in a Monte Carlo tool including NLO QCD corrections will be presented in a subsequent publication.

\section{Higgs PO in VBF production}
\label{sect:VBF}

\subsection{VBF kinematics}
\label{sec:vbf_kinematics}
 Vector boson fusion Higgs production is the largest of all electroweak Higgs production mechanisms in the SM at the LHC. It is 
  highly relevant in the context of experimental Higgs searches due to its striking signature, i.e.~two highly energetic forward jets in  opposite detector hemispheres, 
  which allows an effective separation from the backgrounds. In this chapter we study the phenomenology of VBF production in the PO framework. 
We mainly concentrate our discussion on measuring the quark contact term PO, $\epsilon_{Z q_i}$ and $\epsilon_{Wu_i d_j}$, namely 
the residues of the single pole terms in the expansion of the longitudinal form factors in  Eq.~(\ref{eq:FLGL}).  

At the parton level (i.e.~in the $q q \to h q q$  hard scattering) the ideal observable  relevant to extract the momentum dependence of the 
factor factors would be the double differential cross section $d^2 \sigma/ d q_1^2 d q_2^2$, where $q_1=p_1-p_3$ and $q_2=p_2-p_4$ are the momenta of the two fermion currents entering the process (here $p_1$, $p_2$ ($p_3$, $p_4$) are the momenta of the initial (final) state quarks). 
These $q^2_{i}$ are also the key variables to test and control the momentum expansion at the basis of the PO decomposition.

\begin{figure}
  \begin{center}
    \includegraphics[width=0.45\textwidth]{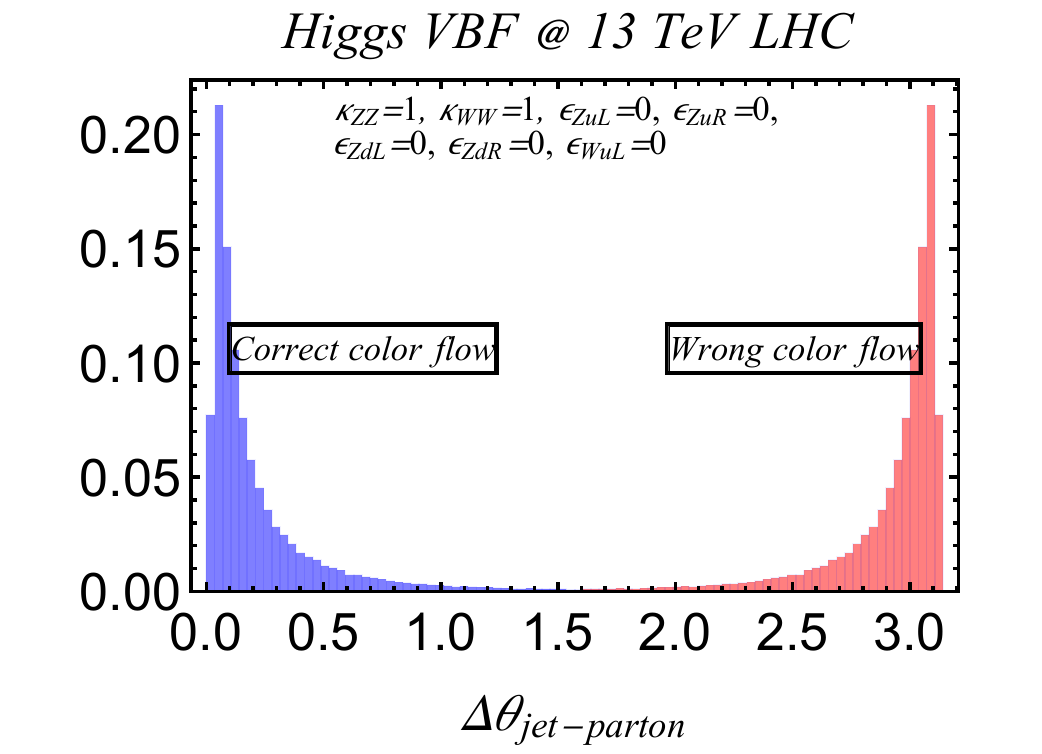}
        \includegraphics[width=0.45\textwidth]{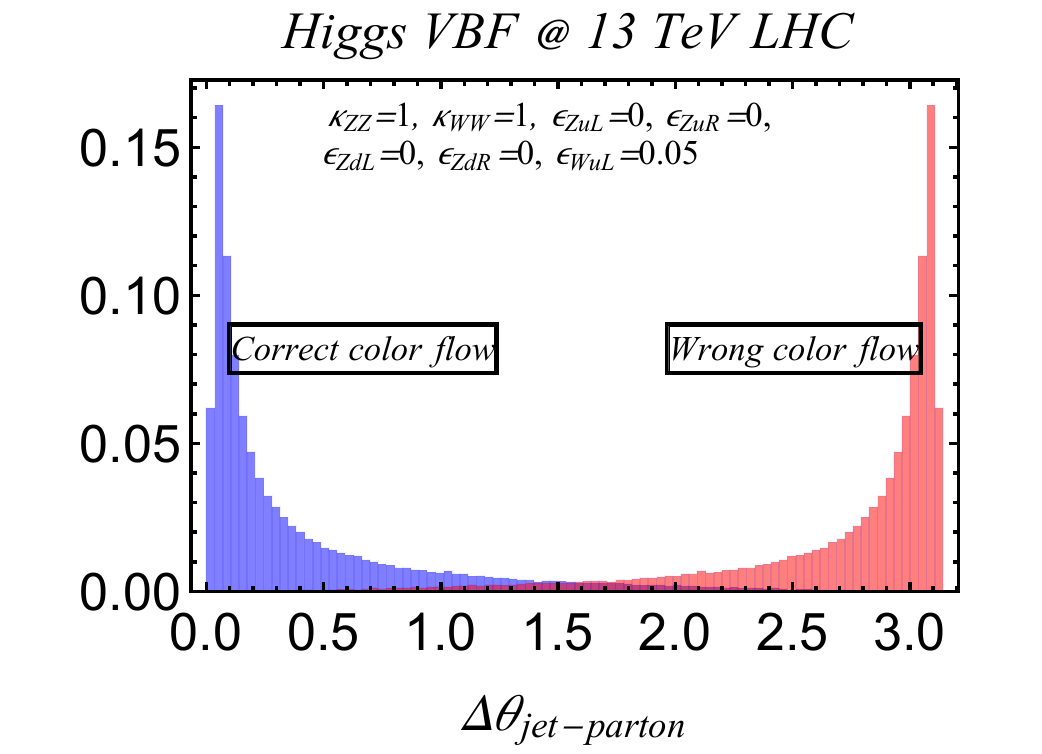}
  \end{center}
\caption{\small\label{fig:colorVSkinemtics} Leading order parton level simulation of the Higgs VBF production at $13$~TeV pp c.m. energy. Show in blue is the distribution in the opening angle of the color connected incoming and outgoing quarks $\measuredangle (\vec p_3,\vec p_1)$, while in red is the distribution for the opposite pairing, $ \angle (\vec p_3,\vec p_2)$. The left plot is for the SM, while the plot on the right is for a specific NP benchmark.   }
\end{figure}

 As a first step of the VBF analysis we have to choose a proper pairing of the incoming and outgoing quarks, given we are experimentally blind to their flavor.
For partonic processes receiving two interfering contributions when the final-state quarks are exchanged, such as $uu \to h u u$ or $ud \to h u d$, the definition of $q_{1,2}$ is even less transparent since a univocal pairing of the momenta can not be assigned, in general, even if one knew the flavor of all partons.
This problem can be  overcome at a practical level by making use of the VBF kinematics, in particular by the fact that the two jets are always very forward. This implies that one can always pair the momentum of the jet going
in one direction with the initial parton going in the same direction. In the same way we can argue  that the interference between different amplitudes (e.g.~neutral current and charged current) is negligible in VBF. In order to check this,
we perform a leading order (LO) parton level simulation of VBF Higgs production ($p p \to h j j$ at $\ord(\alpha^3)$)  employing \MadGraph~\cite{Alwall:2014hca} (version 2.2.3) at $13$~TeV c.m. energy together with the Higgs PO \UFO model. In this simulation we  impose the basic set of VBF cuts,
\begin{align}
\label{eq:vbf_cuts}
p_{\mathrm{T},\mathrm{j}_{1,2}} >30~\GeV,\quad  |\eta_{\mathrm{j}_{1,2}}| <4.5, \quad \mathrm{and}  \quad m_{\mathrm{j}_1\mathrm{j}_2}>500~\GeV.
 \end{align}
   In Fig.~\ref{fig:colorVSkinemtics}, we show the distribution in the opening angle of the incoming and outgoing quark momenta for the two different pairings. The left plot  shows the SM, while the right plot  shows a specific NP benchmark point.  Depicted in blue is the pairing based on the leading color connection using the color flow variable in the event file, while in red  we show the opposite pairing. The plot shows that the momenta of the color connected quarks tend to form a small opening angle and the overlap between the two curves, i.e.~where the interference effects might be sizable, is negligible. This implies that in the experimental analysis the pairing should be done based on this variable.  Importantly, the same conclusions can be drawn in the presence of new physics contributions to the contact terms.

There is a potential caveat to the above argument: the color flow approximation ignores the interference terms that are higher order in $1/N_C$. 
Let us consider a process with two interfering amplitudes with the final state quarks exchanged, for example in $u u \to u u h$. The differential cross section receives three contributions proportional to  $|F^{f f^\prime}_L (t_{13},t_{24})|^2$,  $|F^{f f^\prime}_L (t_{13},t_{24}) F^{f f^\prime}_L (t_{14},t_{23})|$  and $|F^{f f^\prime}_L (t_{14},t_{23})|^2$, where $t_{ij}=(p_i-p_j)^2=-2E_i E_j (1-\cos \theta_{i j})$. For the validity of the momentum expansion it is important that the momentum transfers ($t_{i j}$) remain smaller than the hypothesized scale of new physics. On the other hand, imposing the VBF cuts, the interference terms turn out to depend on one small and one large momentum transfer. However, thanks to the pole structure of the form factors,  
they give a very small contribution.

\begin{figure}
  \begin{center}
    \includegraphics[width=0.48\textwidth]{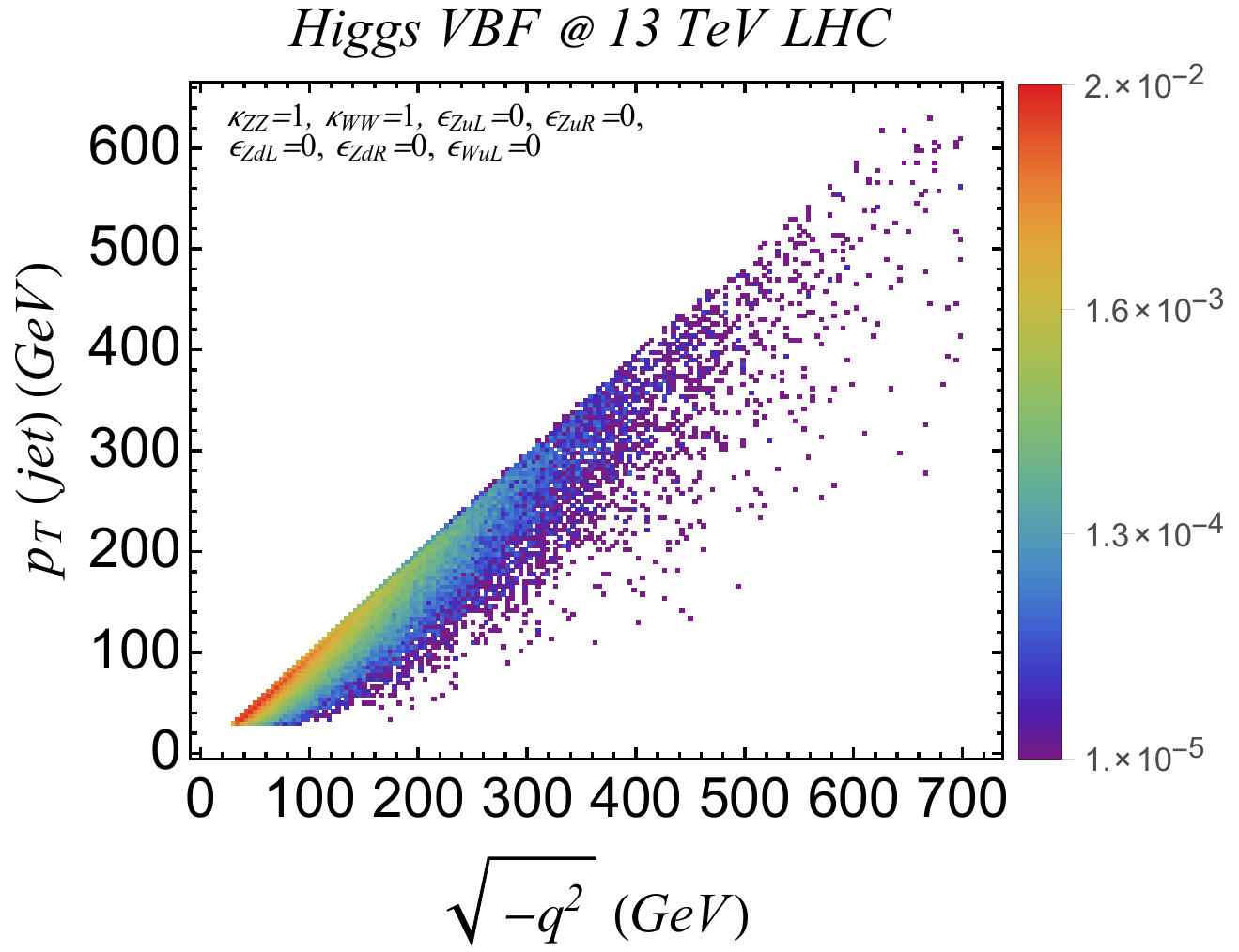}
        \includegraphics[width=0.495\textwidth]{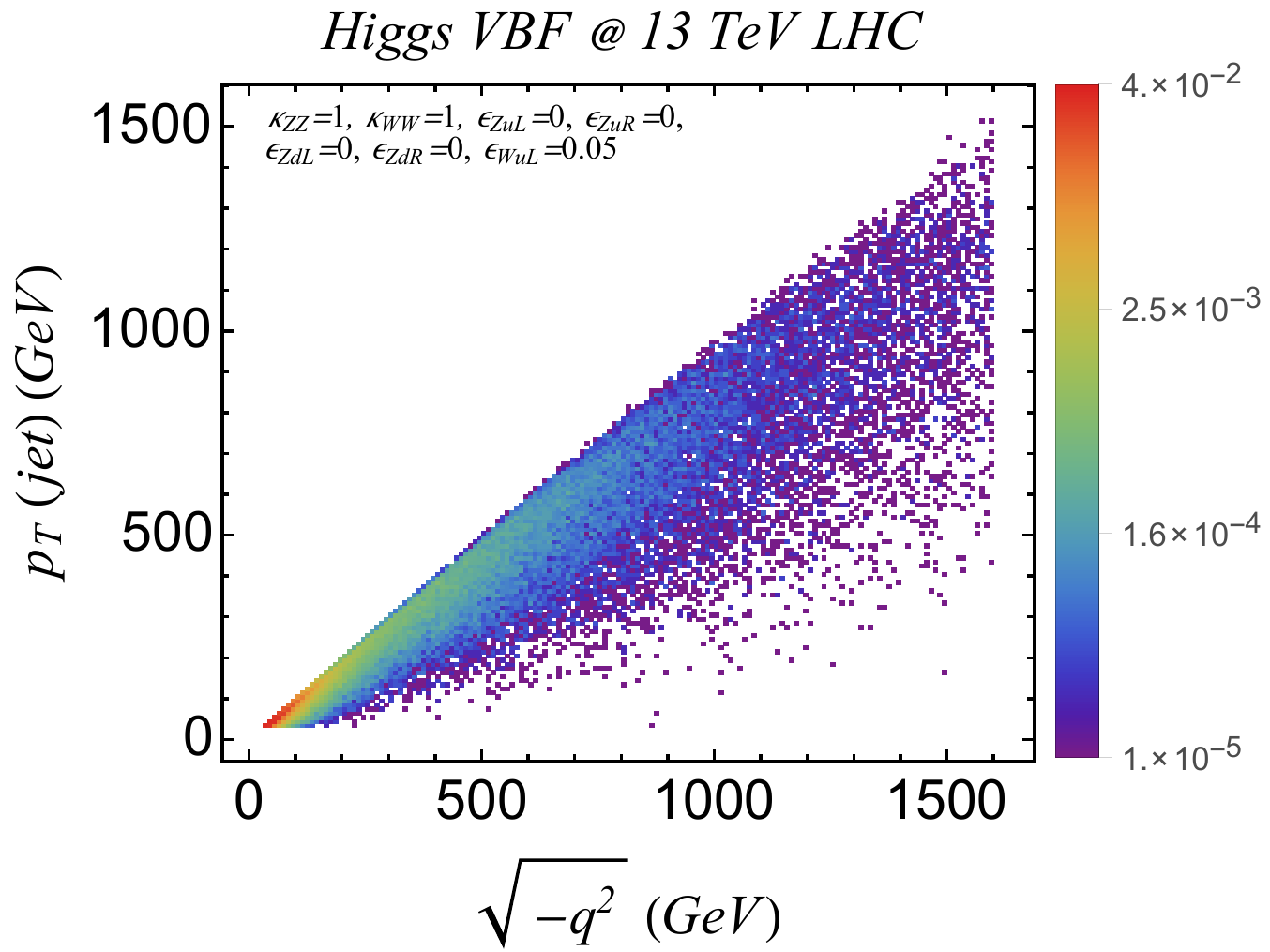}
  \end{center}
\caption{\small\label{fig:pTVSq2} Leading order parton level simulation of the Higgs VBF production at $13$~TeV pp c.m. energy. Shown here is the density histogram in two variables; the outgoing quark $\pT$ and the momentum transfer $\sqrt{- q^2}$ with the initial ``color-connected'' quark. The left plot is for the SM, while the plot on the right is for a specific NP benchmark.      }
\end{figure}

Even though in some  experimental analyses, after reconstructing the momenta of the two VBF tagged jets and the Higgs boson, one could in principle compute the relevant momentum transfers  $q_1$ and $q_2$, adopting the pairing based on the opening angle, in an hadron collider environment like the LHC this is unfeasible.
 Furthermore, for  other Higgs decays modes, such as 
$h\to 2\ell2\nu$, it is not possible to reconstruct the Higgs boson momentum.
Therefore, we want to advocate the use of the $\pT$ of the VBF jets as a proxy for the momentum transfer $q^2_{1,2}$. 
 The quality of this approximation can be understood by explicitly computing the momentum transfer $q^2_{1,2}$ in the VBF limit $|\pT| \ll E_{\mathrm{jet}}$ and for a Higgs produced close to threshold. 
Let us consider the partonic momenta in the c.o.m.~frame for the process:  $p_1 = (E, \vec{0}, E)$, $p_2 = (E, \vec{0}, -E)$, $p_3 = (E'_1, \vec{p}_\mathrm{T,j_1}, \sqrt{E^{\prime 2}_1-\pTjone^2})$ and $p_4 = (E'_2, \vec{p}_\mathrm{T,j_2}, \sqrt{E^{\prime 2}_2 - \pTjtwo^2})$. Conservation of energy for the whole process dictates $2E = E'_1 + E'_2 + E_h$, where $E_h$ is the Higgs energy, usually of order $m_h$ if the Higgs is not strongly boosted. In this case $E - E'_i = \Delta E_i \ll E$ since the process is symmetric  in $1 \leftrightarrow 2$. For each leg, energy and momentum conservation (along the $z$ axis) give
\be
	\left\{ \begin{array}{l} q^z_i = E - \sqrt{E^{\prime 2}_i - p_{\mathrm{T}i}^2}\,, \\ q^0_i = E - E'_i \,,\end{array} \right.\ \quad \to \quad
	\left\{ \begin{array}{l} q^0_i - q^z_i = \sqrt{E^{\prime 2}_i - p_{\mathrm{T}i}^2} - E'_i \approx - \frac{p_{\mathrm{T}i}^2}{2 E'_i} \,, \\
	q^0_i + q^z_i \approx 2 \Delta E_i + \frac{p_{\mathrm{T}i}^2}{2 E'_i} \,. \end{array}\right. ~.
\ee
Putting together these two relations, one finds
\be
	q^2_i \approx - p_{\mathrm{T}i}^2 - \frac{p_{\mathrm{T}i}^2 \Delta E_i}{2 E'_i} + \mathcal{O}(p_{\mathrm{T}i}^4 / E'^2) \approx - p_{\mathrm{T}i}^2 \,,
\ee
where in the last step we assumed $\Delta E_i \ll E'$, i.e. 
the Higgs being produced near threshold.

\begin{figure}
  \begin{center}
   \includegraphics[width=0.47\textwidth]{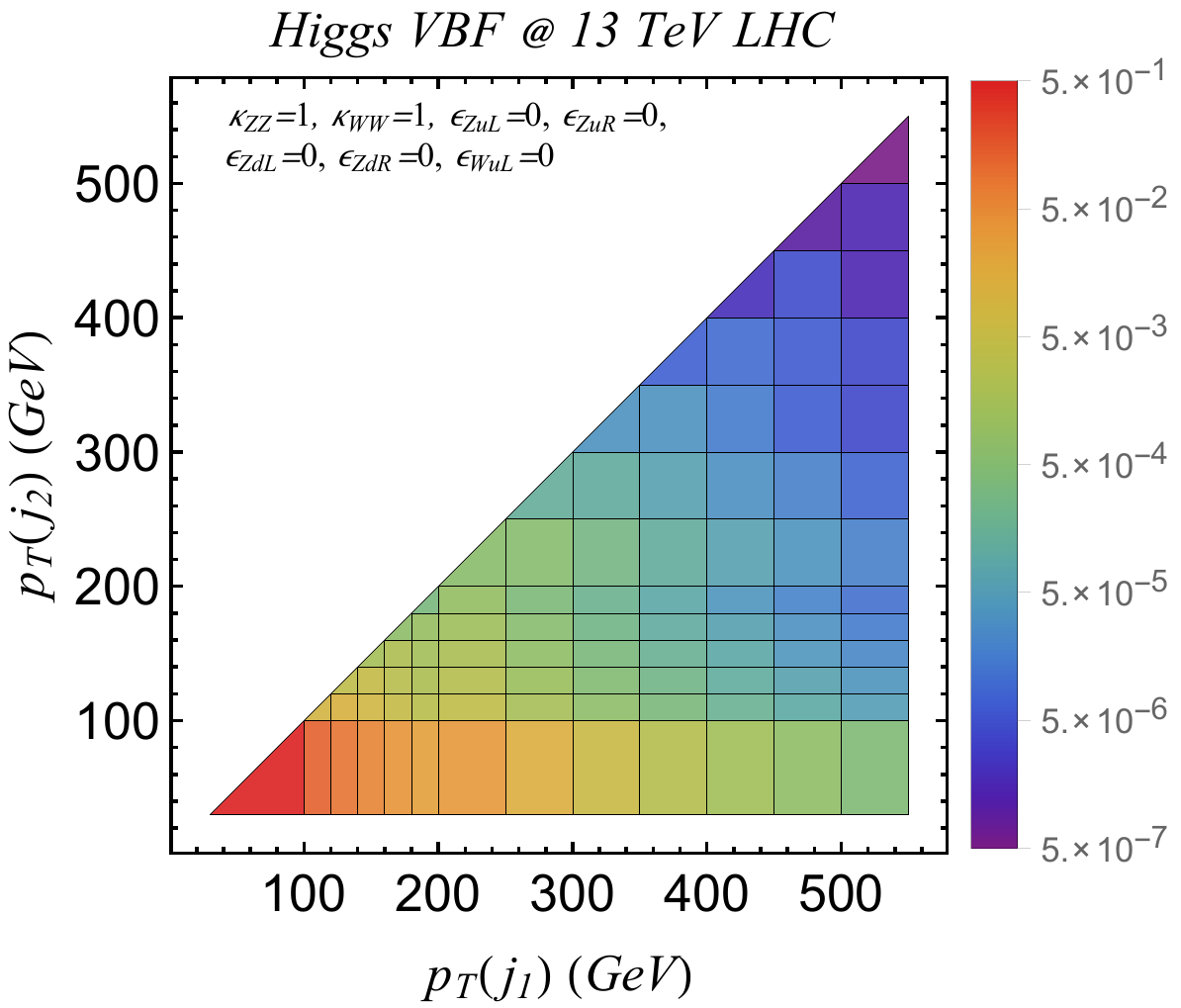}
   \includegraphics[width=0.47\textwidth]{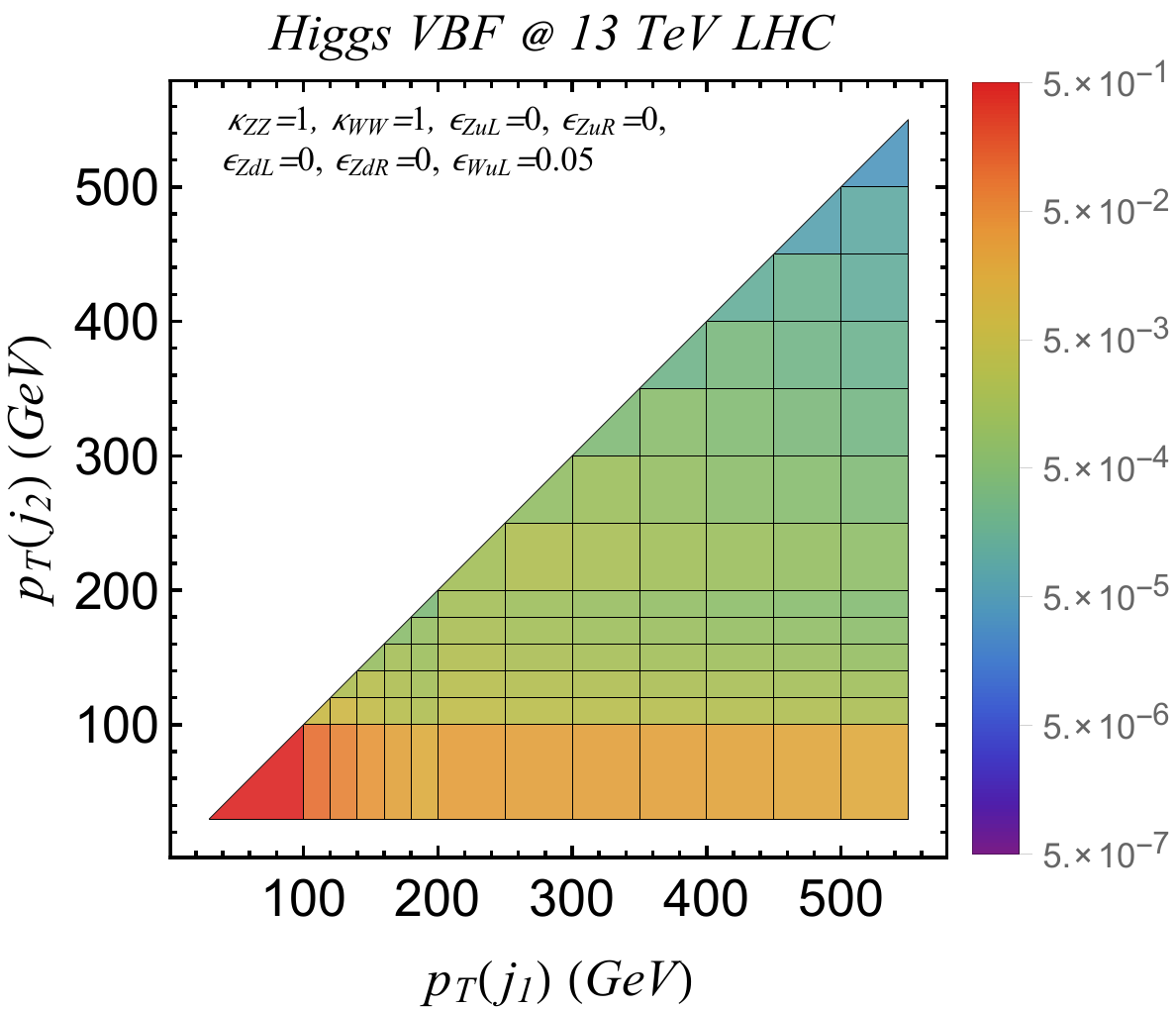}
  \end{center}
\caption{\small\label{fig:smpTpT} Double differential distribution in the two VBF-tagged jet $\pT$ for VBF Higgs production at 13 TeV LHC. The distribution is normalized such that the total sum of events in all bins is 1. (Left) Prediction in the SM. (Right) Prediction for NP in $\epsilon_{W u_L} =0.05$.}
\end{figure}

In order to confirm  the above conclusion, in Fig.~\ref{fig:pTVSq2} we show a density histogram in two variables: the (observable) $\pT$ of the outgoing jet and the
 (unobservable) momentum transfer $\sqrt{-q^2}$ obtained from the correct color flow pairing (the left and the right plots are for the SM and for a specific NP benchmark, respectively). These plots indicate  a very strong correlation of the jet $\pT$ with the momentum transfer $\sqrt{-q^2}$ associated with the correct color pairing. 
We stress that this conclusion holds both within and beyond the SM. 
 Therefore, we encourage the experimental collaborations to report the unfolded measurement of the double differential distributions in the two VBF tagged jet $\pT$'s: $\tilde F(p_{T j_1}, p_{T j_2})$. This measurable distribution is indeed closely related to the form factor entering the amplitude decomposition, $F_L(q_1^2, q_2^2)$, and encode (in a model-independent way) the dynamical information about the high-energy behavior of the process.   
Moreover, as we will discuss in Section~\ref{sec:VBF_prospect}, the extraction of the PO in VBF 
must be done preserving the validity of the momentum expansion: the latter can be checked and enforced setting 
appropriate  upper cuts on the $\pT$ distribution. As an example of the strong sensitivity of the (normalized) $\tilde F(p_{T j_1}, p_{T j_2})$ distribution to NP effects, in Fig.~\ref{fig:smpTpT}, we show the corresponding prediction in the SM (left plot) and
 for a specific NP benchmark (right plot).

\subsection{NLO QCD corrections in VBF}
\label{sec:vbf_nlo}

Inclusive VBF Higgs production in the SM is very stable with respect to higher order QCD corrections \cite{Han:1992hr,Figy:2003nv,Dittmaier:2011ti,Dittmaier:2012vm}. Employing a fixed renormalization and factorization scale $\mu_{\rR,\rF}=m_W$ inclusive NLO QCD corrections are at the level of $5-10\%$ with remaining scale uncertainties of a few percent. At the NNLO QCD level these uncertainties on the inclusive cross section are further reduced below $1\%$~\cite{Bolzoni:2010xr,Bolzoni:2011cu}. 
However, in more exclusive observables, like the $\pT$ spectra of the VBF jets, or when more exclusive experimental selection cuts are applied, sensitivity to QCD radiation is more severe~\cite{Figy:2003nv}, yielding  non-negligible NLO correction factors while NLO scale uncertainties remain small (mostly well below 10\%). Recently the dominant NNLO QCD corrections have been calculated fully differentially~\cite{Cacciari:2015jma} pointing towards a non-trivial phase-space dependence with $5-10\%$ corrections with respect to NLO. 
Besides higher-order corrections of QCD origin, also EW corrections are relevant for VBF Higgs production~\cite{Ciccolini:2007jr,Ciccolini:2007ec}. At an inclusive level they amount to about $-5\%$~\cite{Ciccolini:2007jr}, while at the differential level due to the presence of large EW Sudakov logarithms they reach for example $-15\%$ for $\pTjone=400$~GeV and $-10\%$ for $\pTjtwo=150$~GeV~\cite{Ciccolini:2007ec}.

In the following we will illustrate that the perturbative convergence for exclusive VBF observables can be improved when using a dynamical scale $\mu_0=\HThalf$ (with $H_{\mathrm{T}}$ being the scalar sum of the $\pT$ of all final state particles) with respect to a fixed scale $\mu_0=m_W$. 
In particular, here we will focus on the $\pT$ spectra of the VBF jets -- as inputs for a fit of the Higgs PO. To this end we employ the fully automated \SherpaOpenLoops framework \cite{Gleisberg:2007md,Gleisberg:2008ta,Cascioli:2011va,OLhepforge,Ossola:2007ax,vanHameren:2010cp} for the simulation of EW production of $pp\to hjj$ at LO and NLO QCD in the SM. 
Before applying the VBF selection cuts defined in Eq.~\ref{eq:vbf_cuts} we cluster all final state partons into anti-$\kT$ jets with $R=0.4$ and  additionally require a rapidity separation of the two hardest jets of $\Delta\eta_{j_1j_2} > 3$. This additional requirement, could slightly reduce the capability of differentiating different tensor structures~\cite{Maltoni:2013sma}, however, such a cut is, on the one hand, experimentally required in order to suppress QCD backgrounds.\footnote{~In fact, in most VBF analyses an even tighter selection of $\Delta\eta_{j_1j_2} > 4.5$ is imposed.}  On the other hand, without such a cut NLO predictions for the $\pT$ spectra of the jets become highly unstable when the VBF jet selection is just based on the hardness of the jets, i.e.~a bremsstrahlung jet is easily amongst the two hardest jets and spoils the correlation between the \pT of the jets and the momentum transfer, as discussed in Section~\ref{sec:vbf_kinematics}.

\begin{figure}[t]
   \centering
    \includegraphics[width=0.40\textwidth]{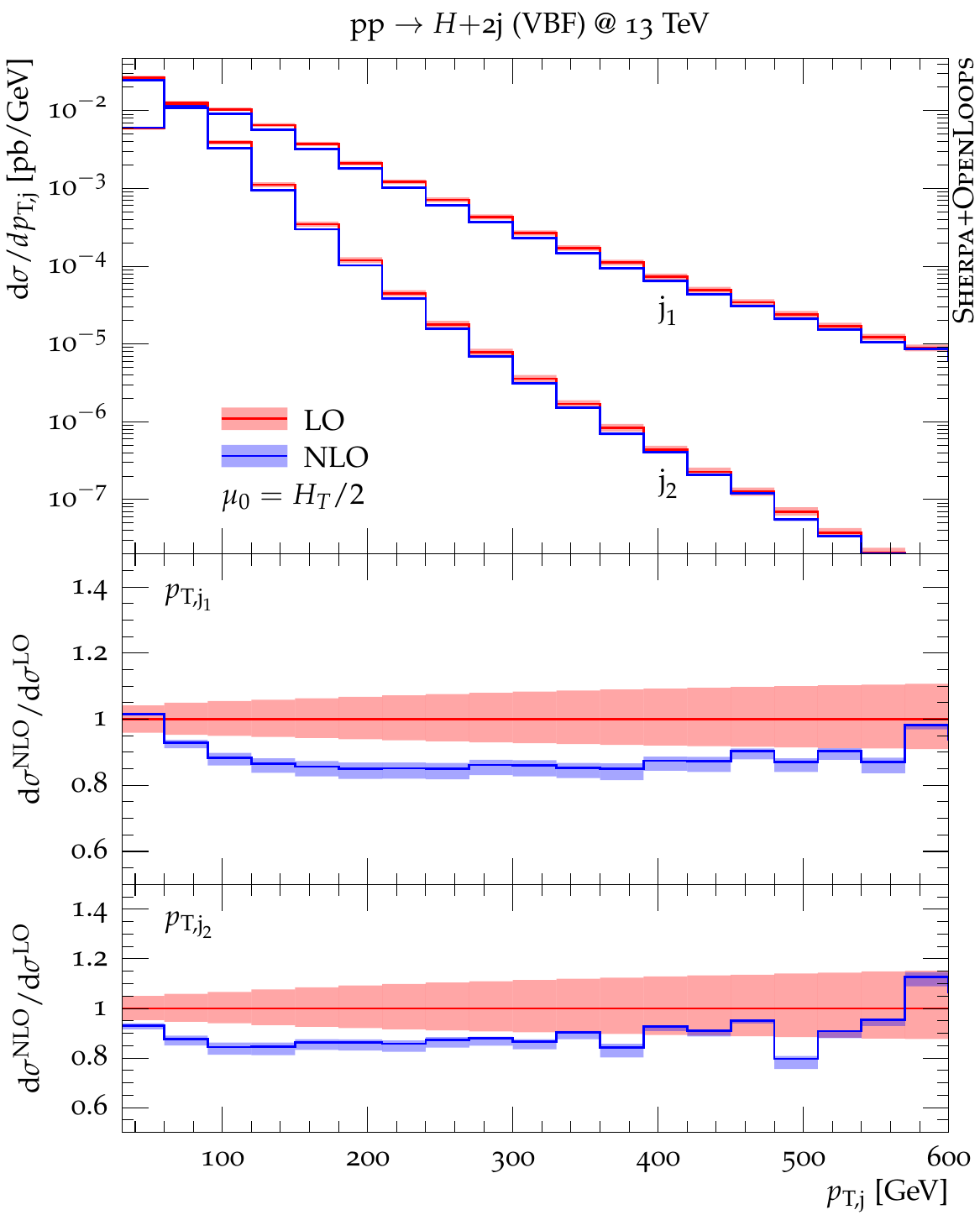}\quad 
       \includegraphics[width=0.48\textwidth]{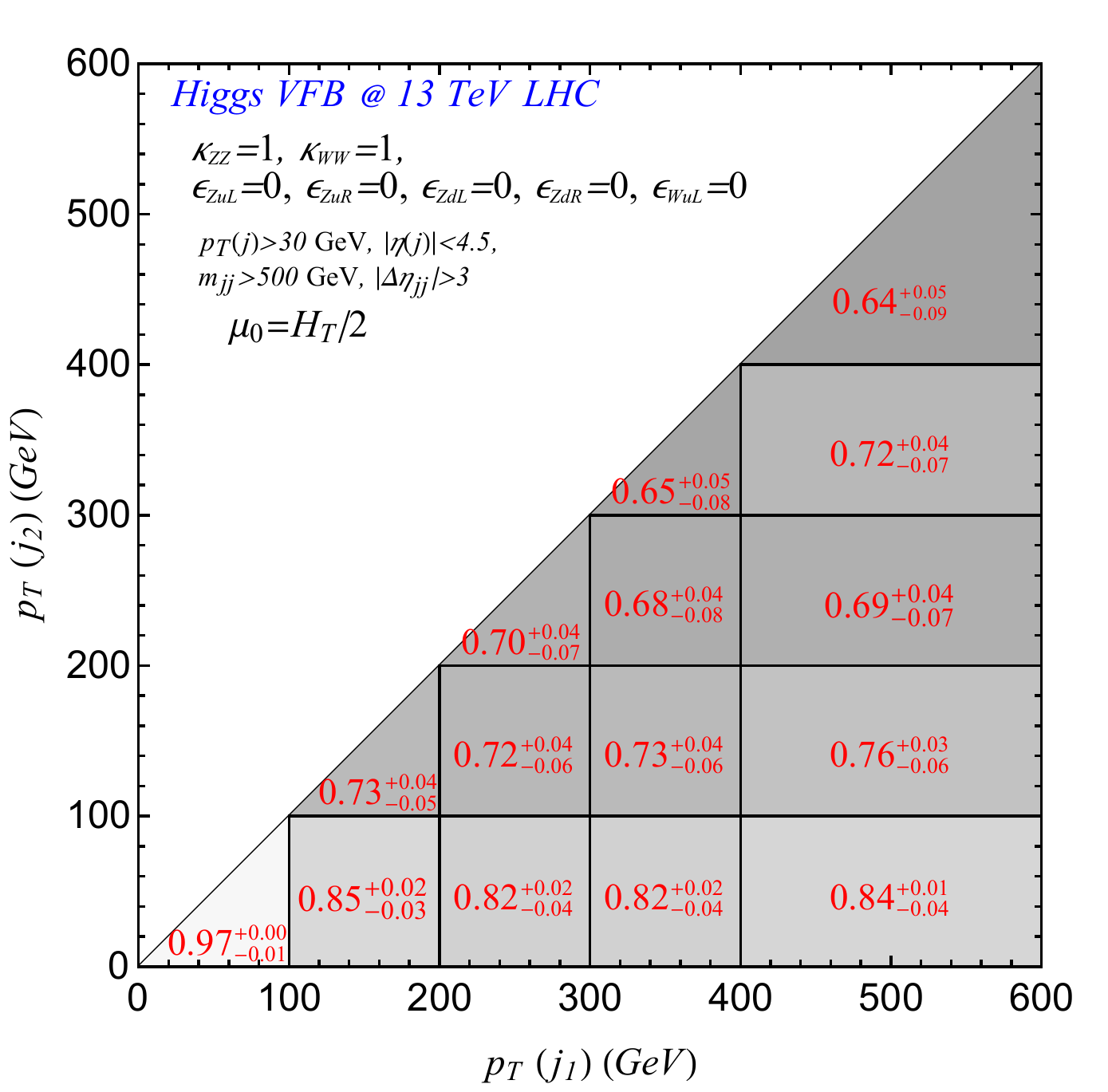}
\caption{One- (left) and two-dimensional (right) NLO correction factors and scale uncertainties for EW production of $pp\to h+2$\,jets in the SM
in function of $\pTjone$ and $\pTjtwo$ employing a central scale $\mu_0=\HThalf$.
}
\label{fig:nlo_kfac_pt_pt}
\end{figure}

In Fig.~\ref{fig:nlo_kfac_pt_pt} we plot the \pT distributions of the hardest and the second hardest jet using a dynamical scale $\mu_0=\HThalf$. On the left one-dimensional \pT spectra are plotted, while on the right we show the corresponding two-dimensional NLO correction factors $K^{\textrm{NLO}}=\sigma^{\textrm{NLO}}/\sigma^{\textrm{LO}}$. 
Here CT10nlo PDFs~\cite{Lai:2010vv} are used both at LO and NLO and uncertainty bands correspond to 7-point renormalization (only relevant at NLO) and factorization scale variations $\mu_{\rR,\rF}=\xi_{R,F}\mu_0$ with $(\xi_\rR,\xi_\rF)=(2,2)$,
$(2,1)$, $(1,2)$, $(1,1)$, $(1,0.5)$, $(0.5,1)$, $(0.5,0.5)$. 
Thanks to the dynamical scale choice NLO corrections to the one-dimensional distributions are almost flat and amount to about $-15\%$, while the dependence in the two dimensional distribution remains moderate with largest corrections for  $\pTjone \approx \pTjtwo$.

In the following section we will detail a fit of Higgs PO based on LO predictions of VBF using the scale choice and setup developed in this chapter. Here we already note, that this fit is hardly affected by the overall normalization of the predictions. Thus, with respect to possible small deviations from the SM due to effective form factor contributions we expect a very limited sensitivity to QCD effects assuming a similar stabilization of higher order corrections as observed for the SM employing the scale choice $\mu_0=\HThalf$.
In order to verify this assumption and to improve on the Higgs PO fit, we are currently extending the simulations within the Higgs PO framework to the NLO QCD level. To this end, the framework has been implemented in the \OpenLoops one-loop amplitude generator in a process independent way. Here, the $\ord(\alpha_S)$ rational terms of $R_2$-type required in the numerical calculation of the one-loop amplitudes in \OpenLoops 
 have been obtained generalising the corresponding SM expressions~\cite{Draggiotis:2009yb}.
The implementation of the dipole subtraction and parton-shower matching in the \Sherpa Monte Carlo framework is based on the model independent \UFO interface of Sherpa~\cite{Hoche:2014kca} and is currently being validated.
 
\subsection{Prospects for the Higgs PO in VBF at the HL-LHC}
\label{sec:VBF_prospect}

The extraction of the PO from the double differential distribution
$\tilde F(p_{T j_1}, p_{T j_2})$  has to be done with  care. 
Here we make an attempt to perform such analysis. In the following  
we estimate the sensitivity of the
HL-LHC, operated at $13$~TeV with $3000$~fb$^{-1}$ of data, on measuring the PO
assuming maximal flavor symmetry in a seven dimensional fit to $\kappa_{ZZ}$,
$\kappa_{WW}$, $\epsilon_{Z u_L}$, $\epsilon_{Z u_R}$, $\epsilon_{Z d_L}$,
$\epsilon_{Z d_R}$ and $\epsilon_{W u_L}$. The ATLAS search for $h\to WW^*$
reported in Ref.~\cite{ATLAS:2014aga} considers the VBF-enriched category in
which the detection of two jets consistent with VBF kinematics is required. The
expected yields in this category are reported in Table VII of
Ref.~\cite{ATLAS:2014aga}. After the final selection cuts  at $8$~TeV  with
$20.3$~fb$^{-1}$ of integrated luminosity, the expected number of Higgs VBF
events in the SM is $4.7$ (compared to $5.5$ background events) in the $e\mu$
sample. Rescaling the number of expected events with the expected HL-LHC luminosity
(3000~fb$^{-1}$) and cross section, we expect about $2000$ SM Higgs VBF events 
to be collected by
 each experiment. In the following, we make a  brave
approximation and neglect  any background events in the fit and assume that the HL-LHC will observe a total of $2000$ events compatible with the SM expectations.

\begin{figure}[t]
   \centering
    \includegraphics[width=0.5\textwidth]{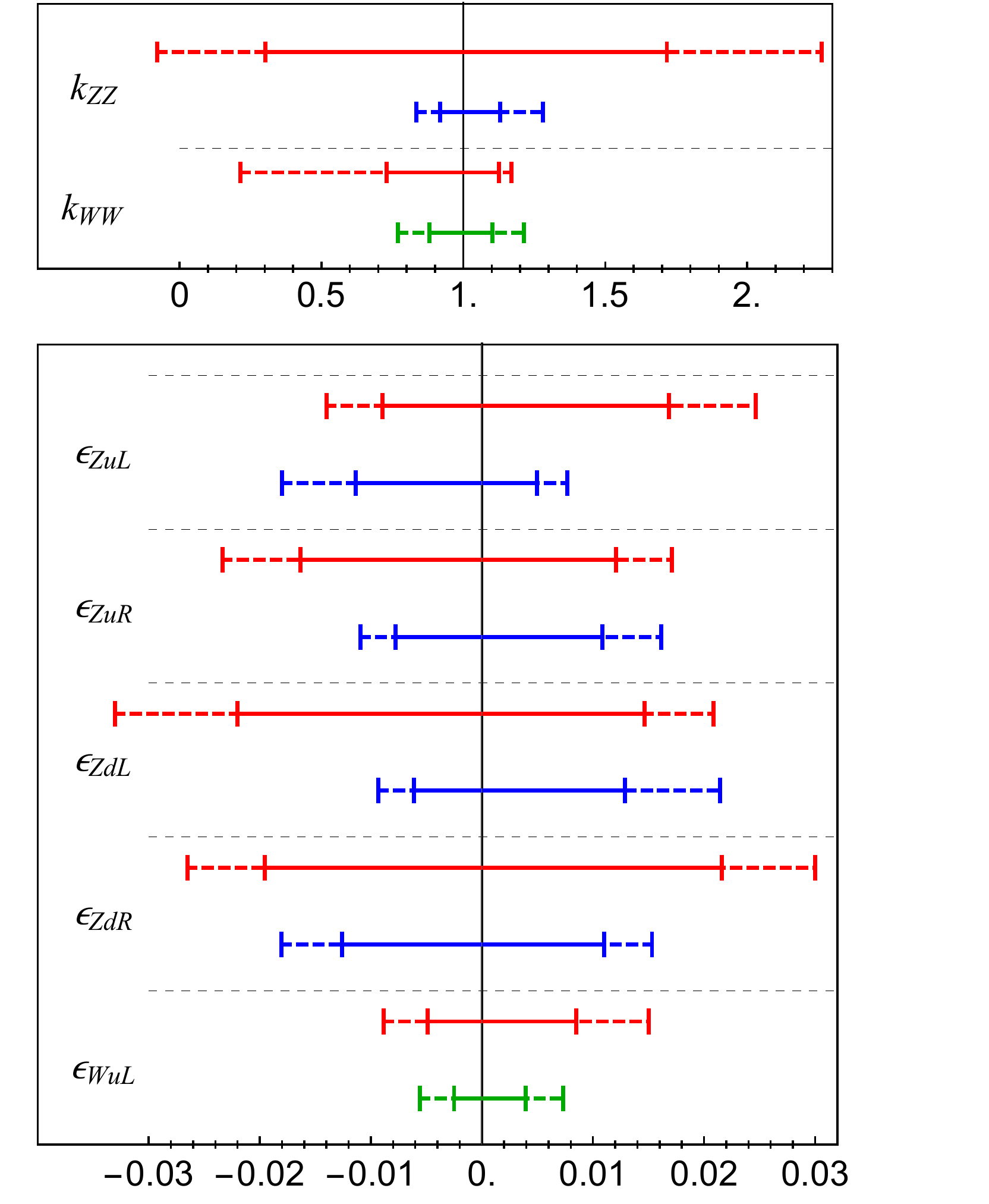}
\caption{\small\label{fig:HPO_HLLHC} Prospects for measuring Higgs PO in electroweak Higgs production at the HL-LHC at 13 TeV with $3000~ \text{fb}^{-1}$ of integrated luminosity. For VBF and $Zh$ we considered the $h \to 2\ell2\nu$ channel (with $Z\to 2\ell$ in $Zh$) while for $Wh$ we considered only the clean $h\to 4\ell$, $W \to \ell \nu$ channel. The solid (dashed) intervals represent the $1\sigma$ ($2\sigma$) constraints in each PO, where all the others are profiled. The red bounds are from VBF, the blue ones  from $Zh$ and the green ones  from $Wh$ production. More details  can be found in the main text.}
\end{figure}

As anticipated, a key point to be addressed for a consistent extraction of the PO is the
validity of the momentum expansion.
In order to control such expansion, we set an upper cut on the  $\pT$ of the leading VBF-tagged jet.  
The momentum expansion of the form factors in Eq.~\eqref{eq:FLGL} only makes sense  if the higher order terms in $q_{1,2}^2$ are suppressed. 
This requirement leads to  the consistency condition,
\be
\epsilon_{X_f} ~ |q_{\rm{max}}^2| \lesssim m_Z^2 ~ g_X^{f}~,
\label{eq:consistency}
\ee
where $q_{\rm{max}}^2$ is the largest momentum transfer in the process. A priori we do not know  the size of  $\epsilon_{X_f}$
or, equivalently, the effective scale of new physics. However, a posteriori we can verify by means of  Eq.~(\ref{eq:consistency}) 
if we are allowed to truncate the momentum expansion to the first non-trivial terms. In practice, 
setting a cut-off on  $\pT$ we implicitly define a value of $\sqrt{-q^2_{\rm{max}}}$. Extracting the $\epsilon_{X_f}$ for 
 $\pTj < (\pTj)^{\rm max} \approx  \sqrt{-q^2_{\rm{max}}}$ we can 
check if Eq.~(\ref{eq:consistency}) is satisfied. Ideally, the experimental collaborations should perform the extraction of the $\epsilon_{X_f}$ 
for different values of $(\pTj)^{\rm max}$ optimizing the range according to the results obtained. 
In the following exercise we set $(\pTj)^{\rm max}  = 600~ \GeV$  which, a posteriori, will turn out to be a good choice in absence of 
any sizeable deviations from the SM.

In our analysis we choose the binning in the double differential distributions in the two VBF tagged jet $\pT$'s as $\{30-100-200-300-400-600\}$~\GeV. We use the 
UFO implementation of the Higgs PO in the Sherpa Monte Carlo generator \cite{Gleisberg:2008ta,Hoche:2014kca} to simulate VBF Higgs events over the relevant PO parameter space in proton-proton collisions at $13$~TeV c.m. energy. Here we employ the VBF selection cuts as listed in Eq.~(\ref{eq:vbf_cuts})
with the additional requirement $\Delta\eta_{j_1j_2} > 3$.  We verified that the results of the fit are independent on the precise value of this last cut. Renormalization and factorization scales are set to $\mu_{\rR/\rF}=\HThalf$, as discussed in Section~\ref{sec:vbf_nlo}.

Analyzing the simulation output,  we find  expressions for the number of expected events in each bin as a quadratic polynomial in the PO:
\begin{equation}
\label{eq:Nevbin}
	N^{\rm ev}_a = \kappa^T X^a \kappa~, \quad \mathrm{with} \quad
	\kappa \equiv (\kappa_{ZZ},\kappa_{WW},\epsilon_{Z u_L}, \epsilon_{Z u_R}, \epsilon_{Z d_L}, \epsilon_{Z d_R},\epsilon_{W u_L})^T~,
\end{equation}
 where $a$ is a label for each bin.
Assuming that the HL-LHC ``would-be-measured'' distribution is SM-like and describing the number of events in each bin with a Poisson distribution, we construct a global likelihood $L$ and evaluate the best-fit point from the maximum of the likelihood. We then define the test statistic, $\Delta \chi^2 = - 2 \log (L/L_{\mathrm{max}})$, as a function of the seven PO. For more details on the statistical analysis see App.~\ref{app:stat}.
\begin{figure}
  \begin{center}
	\includegraphics[width=0.47\textwidth]{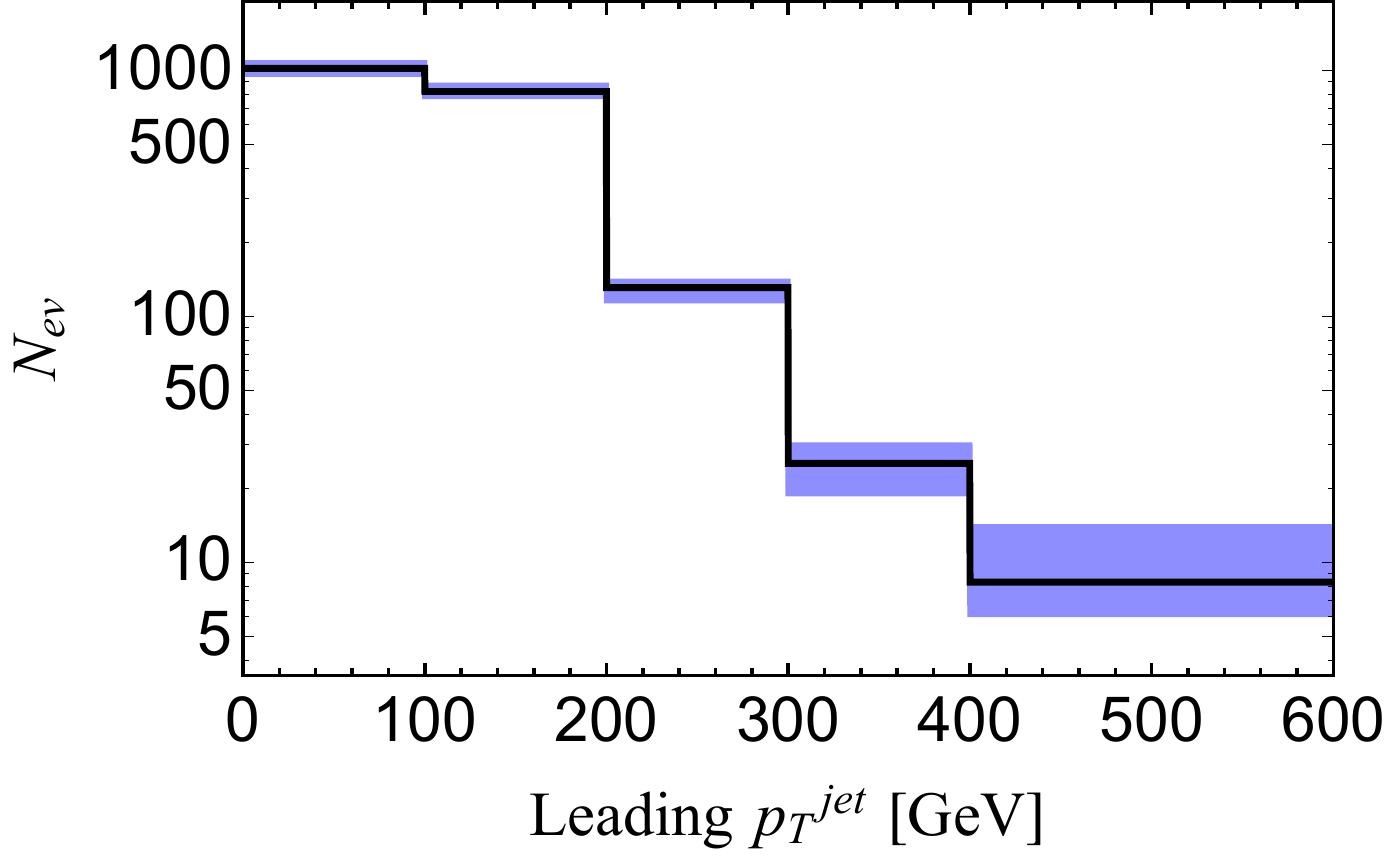} \quad
        \includegraphics[width=0.48\textwidth]{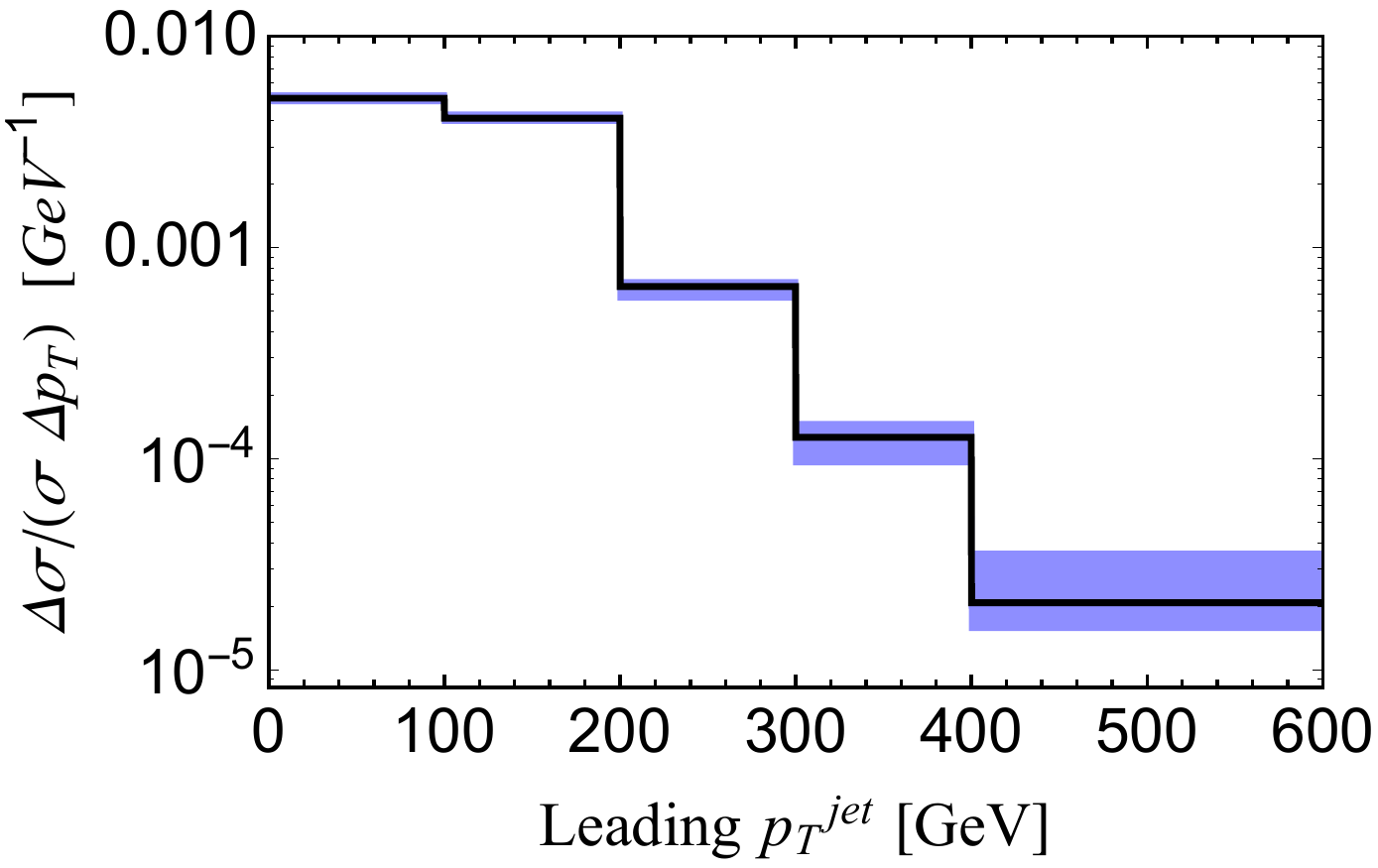}
  \end{center}
\caption{\small\label{fig:VBF_lead_pTj_hist} Allowed deviations in the distribution of the leading-jet $\pT$ by varying the PO within the $-2\log L / L_{\rm max} < 4$ ($2\sigma$) region obtained after the VBF fit. In the left plot we show the absolute number of events in each bin, while in the right one we show the normalized distribution with respect to the total number of events and the bin width.}
\end{figure}

In Fig.~\ref{fig:HPO_HLLHC}, we show in red the $1\sigma$ ($\Delta \chi^2 \le 1$) and $2\sigma$ ($\Delta \chi^2 \le 4$) bounds for each PO, while profiling over all the others. The expected uncertainty on the $\kappa_{ZZ,WW}$ is rather large (with  a loosely bounded direction: $\delta \kappa_{ZZ} \approx -3 \delta\kappa_{WW}$), however in a global fit to all Higgs data, these PO are expected to be much more precisely constrained from  $h\to 4\ell, 2\ell 2\nu$ decays. The  most important  
conclusion of this analysis is that at the HL-LHC all  five  production PO can be constrained at the percent level. 
In the following we test the robustness of this conclusion.

The likelihood obtained from the PO fit is highly non-Gaussian, which is mainly due to the fact that Eq.~\eqref{eq:Nevbin} is quadratic in the PO, and thus the $\Delta \chi^2$ is approximately a quartic polynomial. This implies that using the Gaussian approximation to obtain the $1\sigma$ uncertainties from an expansion around the minimum overestimates these errors (compare with the $1\sigma$ intervals of Fig.~\ref{fig:HPO_HLLHC}):
\be\begin{split}
	{\rm VBF:} \qquad \sigma^{\rm Gauss}_{\rm quad}(\kappa_{ZZ},  \kappa_{WW}, \epsilon_{Z u_L}, \epsilon_{Z u_R}, \epsilon_{Z d_L}, \epsilon_{Z d_R}, \epsilon_{W u_L}) =& \\
	= (0.63, 0.18, 0.021, 0.026, 0.032, 0.050, 0.008)~. &
	\label{eq:VBF_GaussFit_Quadr}
\end{split}
\ee
In order to assess if  these bounds simply come from the information of the total rate, which in a complete analysis depends 
 on the decay parameters and the  total Higgs decay width, or it  indeed stems from the shape analysis, we introduce a new parameter $\mu$ as an overall rescaling of the number of events in all bins, $N^{\rm ev}_a \to \mu N^{\rm ev}_a$. We then perform the same fit as above with this extra parameter and subsequently profile over it.\footnote{~In order to stabilize the fit we assign a Gaussian distribution for $\mu$ centered around 1  with  $\sigma = 10$.}  As a result, $\kappa_{ZZ}$ and $\kappa_{WW}$ become unconstrained but the constraints on the contact terms do not change qualitatively. We thus conclude that their bounds do come from the
 shape information, i.e. the normalized distribution $\tilde F(p_{T j_1}, p_{T j_2})$.

Furthermore,  we have checked that the uncertainties on the entries of the $X^a$ matrices, due to the finite statistics of our Monte Carlo simulations, do not impact  the fit results. Details of this analysis are reported in App.~\ref{app:stat}. The approach sketched there can also be used to estimate the uncertainty of our result caused by missing higher order theory corrections, most notably NLO electroweak effects. As anticipated, 
the latter  can exceed the 10\% level in VBF~\cite{Ciccolini:2007jr,Ciccolini:2007ec}; however,  the largest contributions are due to factorizable corrections (EW Sudakov logarithms and soft QED radiation) 
that can be reabsorbed by a redefinition of the PO. From the results in Ref.~\cite{Denner:2003iy} for the related process $e^+ e^- \to \nu \bar\nu h$ we estimate non-factorizable NLO electroweak corrections to barely reach $10\%$ in some dedicated corners of the phase space (being typically well below such values in most of the phase space). To be conservative, we assign uncorrelated relative errors of $10\%$ in each element of the matrices $X^a$, by introducing  appropriate nuisance parameters, and redo the fit. Profiling over these nuisance parameters, in the Gaussian approximation, we find the following $1\sigma$  uncertainties for the PO:  $\Delta \kappa_{ZZ} = 0.94$, $\Delta \kappa_{WW} = 0.31$, $\Delta \epsilon_{Z u_L} = 0.022$, $\Delta \epsilon_{Z u_R} = 0.027$, $\Delta \epsilon_{Z d_L} = 0.033$, $\Delta \epsilon_{Z d_R} = 0.055$ and $\Delta \epsilon_{W u_L} = 0.009$. Interestingly, comparing these with the Gaussian errors shown above, we conclude that the estimated sensitivity does not worsen significantly, indicating that statistical errors will still dominate. It is worth noting that the theoretical uncertainties are more relevant for 
the determination of $\kappa_{ZZ}$ and $\kappa_{WW}$ and less relevant for the contact terms PO.

Now that we have obtained the constraint on the PO, we can a posteriori check the consistency condition of the analysis, namely, that we are in the regime of small deviations from the SM prediction. In Fig.~\ref{fig:VBF_lead_pTj_hist}, we show the envelope of the allowed deviations in the leading-jet $\pT$ distribution, obtained by varying the PO inside the $2\sigma$ region. As can be seen, the size of the distribution is well constrained  up to $400~\textrm{GeV}$.
Equivalently, using $| \epsilon_{X_f} | \lesssim 0.01$ to check the consistency condition (\ref{eq:consistency}), we find 
 $0.01 \times (600~\textrm{GeV})^2 / m_Z^2 \lesssim 1$, suggesting that we have performed an analysis in a kinematical region 
 where the momentum expansion is indeed reliable. 
 
\section[Higgs PO in \VH~production]{Higgs PO in \VH~production}
\label{sect:VH}

\subsection[VH kinematics]{VH kinematics}

Higgs production in association with a $W$ or $Z$ boson are respectively the third and fourth most important Higgs production processes in the SM, by total cross section. Combined with VBF studies, they offer  complementary  handles to limit and disentangle the various Higgs PO. Due the lower cross sections, so far these processes are mainly studied in the highest-rate Higgs decay channels, such as $h\to b\bar{b}$ \cite{Aad:2012gxa,Chatrchyan:2012ww,Chatrchyan:2013zna,Aad:2014xzb} and $h\to WW^*$ \cite{Chatrchyan:2012qr,Khachatryan:2014jba,Aad:2015gba,Aad:2015ona}. The drawback of these channels  are large backgrounds, which  are overwhelming in the $b\bar{b}$ case and of the same order as the signal in the $WW^*$ channels. In the following we skip over the challenges and the difficulties due to the presence of large backgrounds in these dominant modes, focusing only on $V+h$ decay channels with a good S/B ratio (that should become accessible at the HL-LHC). In those channels we analyze the prospects for the extraction of the 
corresponding production PO.

\begin{figure}
  \begin{center}
    \includegraphics[width=0.45\textwidth]{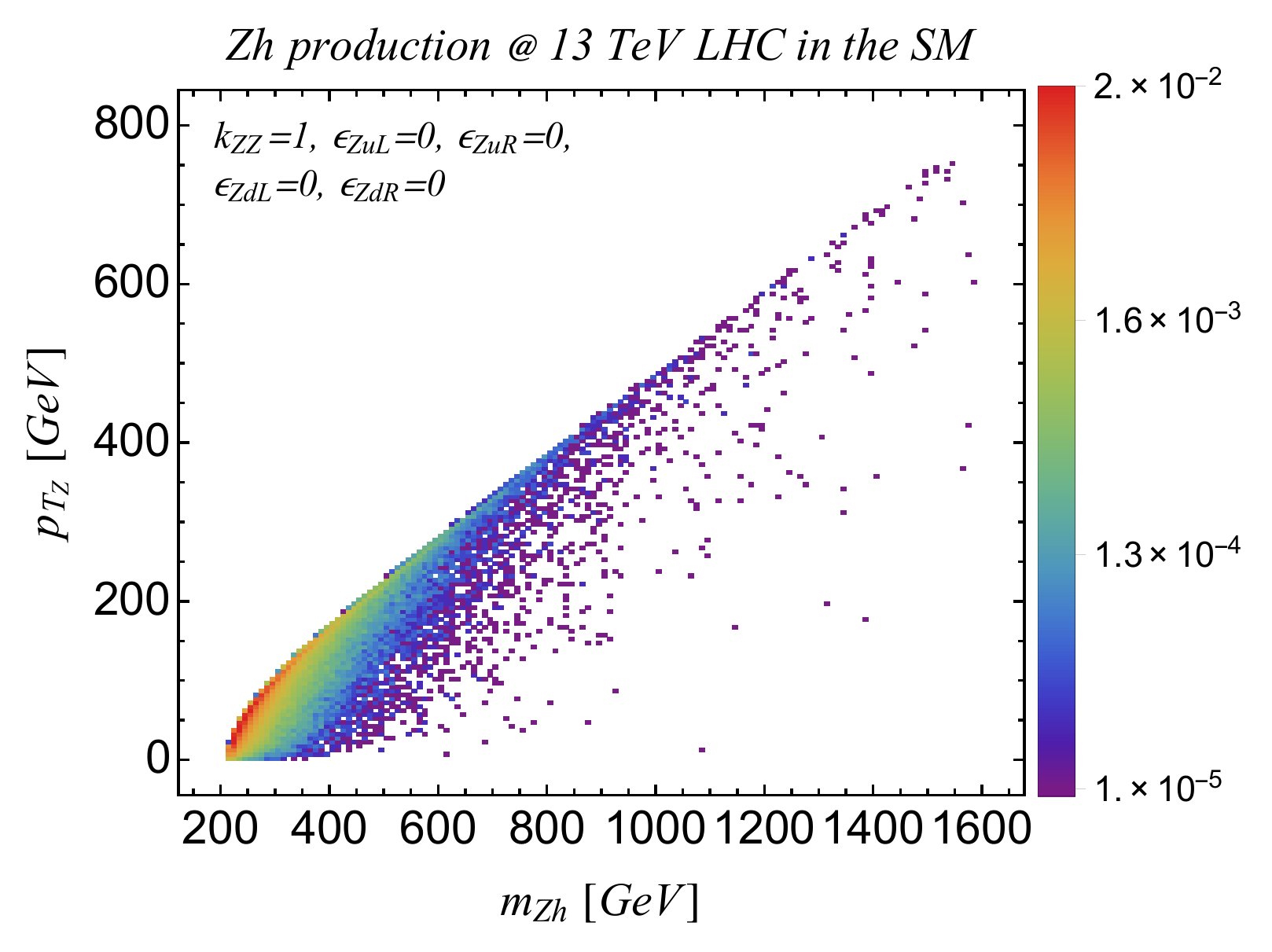} \quad
    \includegraphics[width=0.45\textwidth]{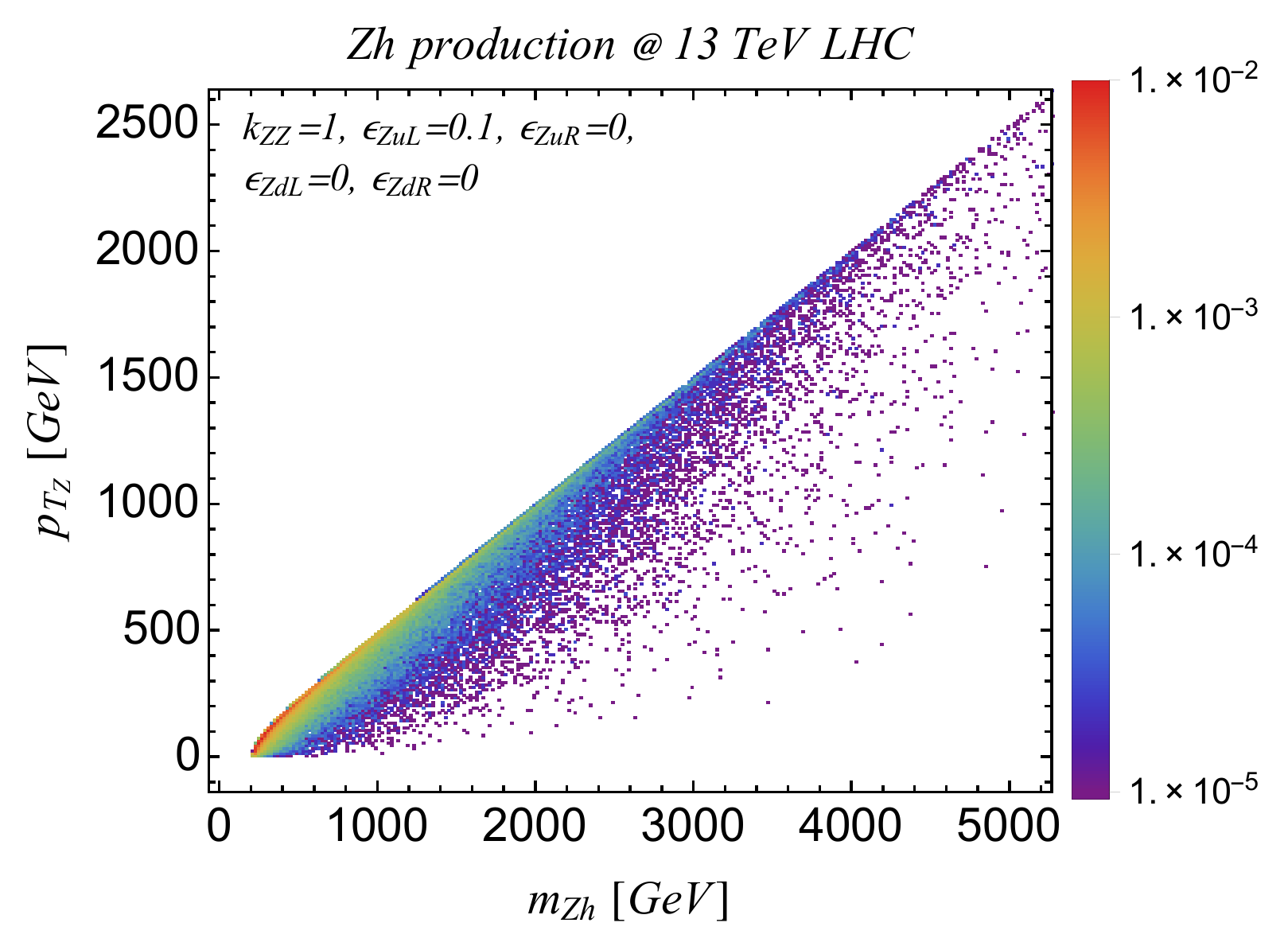}
  \end{center}
\caption{\small\label{fig:pTZ_qSQ_SM} The correlation between the $Zh$ invariant mass and the $\pT$ of the $Z$ boson in $Zh$ associate production at the 13TeV LHC in the SM (left plot) and for a BSM point $\kappa_{ZZ} = 1$, $\epsilon_{Z u_L} = 0.1$ (right plot). A very similar correlation is present in the $Wh$ channel.}
\end{figure}

An important improvement for future studies of these channels with the much higher luminosity that will be available,  can be obtained scrutinizing   differential distributions in  specific kinematical variables. In Section~\ref{sec:VHprod_ff} we showed that with this respect  the (not always measurable) invariant mass of the \Vh system is the most important observable in this process, since the form factors directly depend on it.
In  channels where the invariant mass $m_{Vh}$ can not be reconstructed due to the presence of neutrinos,  an accessible kinematical proxy exhibiting a sizable correlation with $q^2$ is given by the transverse momentum of the vector boson $\pTV$ or, equivalently, that of the Higgs, as can be seen in the Fig.~\ref{fig:pTZ_qSQ_SM}. Even though this correlation is not as good as the one between the jet $\pT$ and the momentum transfer in the VBF Higgs production channel, a measurement of the vector boson (or Higgs) $\pT$ spectrum
would still offer important information on the underlying structure of the form factors appearing in Eq.~\eqref{eq:FFVh}, namely $F_L^{q_i Z}(q^2)$ or $G_L^{q_{ij} W}(q^2)$, see also Ref.~\cite{Englert:2015hrx}.
The invariant mass of the \Vh system is given by  $m_{Vh}^2 = q^2 = m_V^2 + m_h^2 + 2 p_V\cdot p_h$.
In the c.m. frame, we have $p_V = (E_V, \vec{p}_{\mathrm{T}}, p_z)$ and $p_h = (E_h, - \vec{p}_{\mathrm{T}}, - p_z)$ and
\be
	m_{Vh}^2 = m_V^2 + m_h^2 + 2 \pT^2 + 2 p_z^2 + 2\sqrt{m_V^2 + \pT^2 + p_z^2}\sqrt{m_h^2 + \pT^2 + p_z^2} \, \stackrel{|\pT| \to \infty}{\longrightarrow} \, 4 \pT^2~.
\ee
For $p_z = 0$ this equation gives the minimum $q^2$ for a given $\pT$, which can be seen as the left edge of the distributions in Fig.~\ref{fig:pTZ_qSQ_SM}. This is already a valuable information, especially to address the validity of the momentum expansion.
For example the boosted Higgs regime  utilized in  many $b\bar{b}$ analyses implies a potentially dangerous lower cut-off on 
$q^2$: here a bin with $\pT > 300$ GeV implies $\sqrt{q^2} \gtrsim 630$ GeV, which might be a problem for the validity of the momentum expansion.

In the $Wh$ process, for a leptonic $W$ boson decay, the $\pTW$ can not be reconstructed independently of the Higgs decay channel. 
 It is tempting to consider the $\pT$ of the charged lepton from the $W$ decay  as correlated with the $Wh$ invariant mass. 
 However, we checked explicitly that any correlation is washed out by the decay.

\subsection[NLO QCD corrections in \VH]{NLO QCD corrections in \VH}

\begin{figure}[t]
   \centering
    \includegraphics[width=0.45\textwidth]{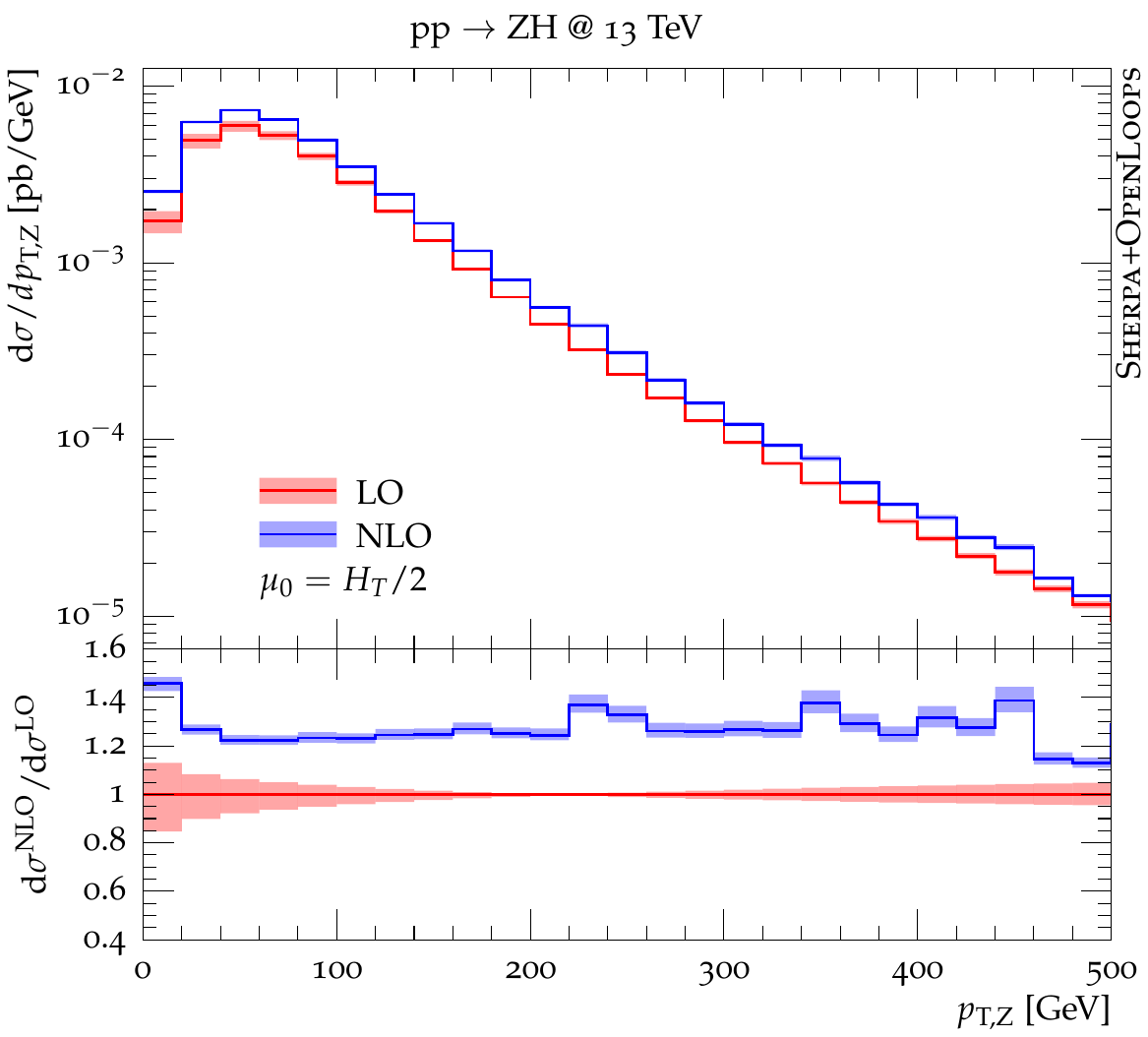}\quad 
       \includegraphics[width=0.45\textwidth]{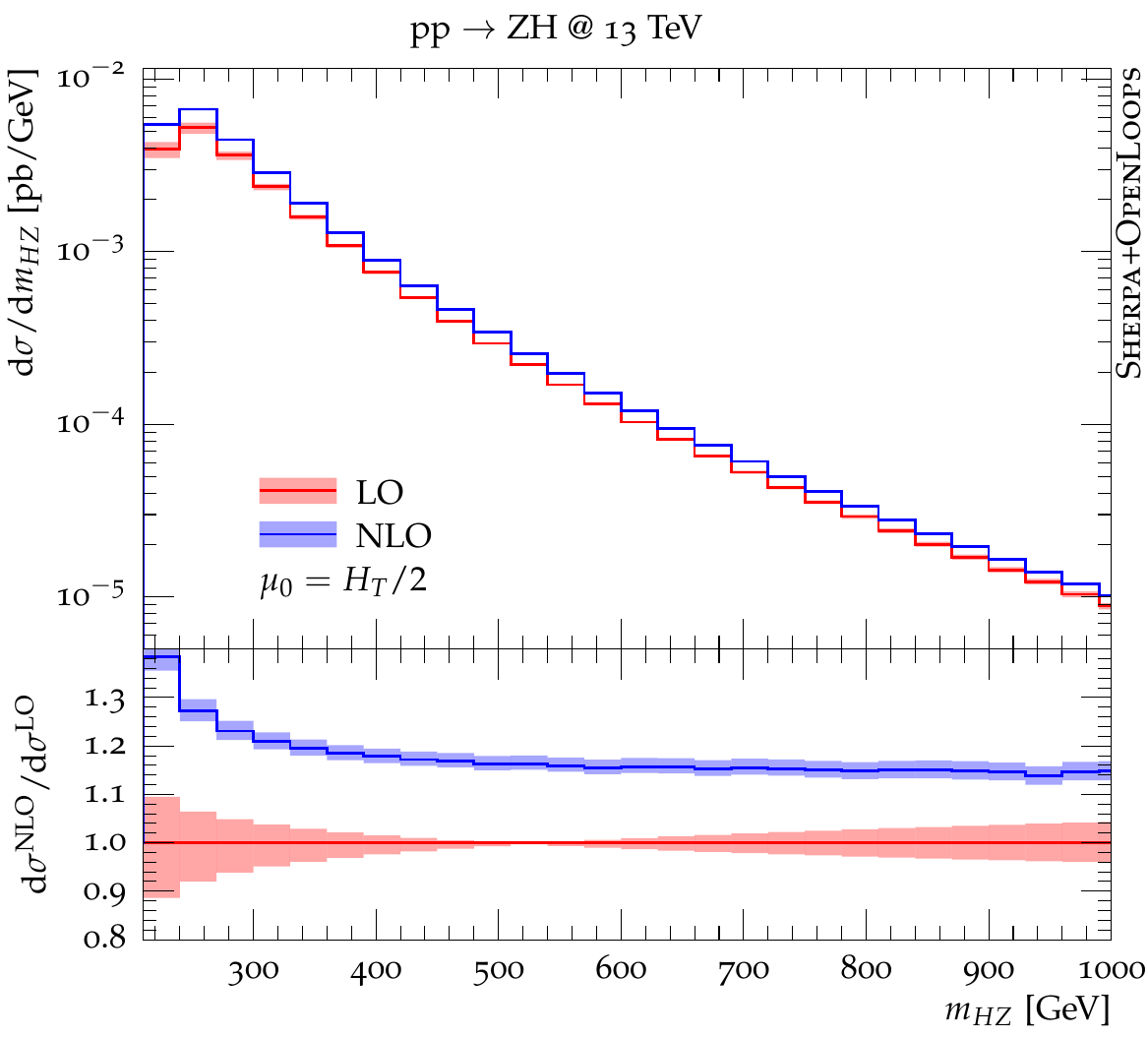}\\
\caption{ NLO correction factors and scale uncertainties for $pp\to ZH$\ in the SM
in function of $\pTZ$ (left) and $\minvZH$ employing a central scale $\mu_0=\HThalf$.
}
\label{fig:nlo_vh_ptZ}
\end{figure}

At the inclusive and exclusive level QCD corrections to \VH~processes are well under control~\cite{Dittmaier:2011ti,Dittmaier:2012vm,Heinemeyer:2013tqa}. The dominant QCD corrections of Drell-Yan-like type are known fully differentially up to NNLO~\cite{Ferrera:2011bk,Ferrera:2013yga,Ferrera:2014lca} and on the inclusive level amount to about $30\%$ with respect to the LO predictions for both $Wh$ and $Zh$. Remaining scale uncertainties are at the level of a few percent. 

In Fig.~\ref{fig:nlo_vh_ptZ} we illustrate the NLO QCD corrections to $Zh$ in the SM looking at differential distributions in $p_{\textrm{T},Z}$ and $m_{Zh}$, while the qualitative picture is very similar for $Wh$. The employed setup is as detailed already in Section~\ref{sec:vbf_nlo}, while here we do not apply any phase-space cuts. Although the natural scale choice for \VH~clearly is $\mu_0=Q= \sqrt{(p_{h}+p_{Z})^2}$, here we employ a scale $\mu_0=\HThalf$. With this scale choice the resulting differential distributions (to be utilized in the Higgs PO fit) are almost free of shape effects due to higher-order QCD corrections. A study of a similar stabilization including deformations in the Higgs PO framework will be performed in the near future.

In the case of $Zh$ besides Drell-Yan-like production there are loop-induced contributions in $g g \to Z h$ mediated by heavy quark loops, which in particular become important in the boosted regime with $\pTH>200$~GeV~\cite{Englert:2013vua,Goncalves:2015mfa}.

Besides QCD corrections also EW corrections give relevant contributions and shape effects to \VH~processes due to Sudakov logarithms at large energies. They are known at NLO EW \cite{Denner:2011id,Denner:2014cla} and decrease the LO predictions by about $10\%$ for  $p_{\textrm{T},Z}=300$~GeV and by about $15\%$ for  $p_{\textrm{T},W}=300$~GeV. We stress that, as in the VBF case, the dominant NLO EW effects are 
factorizable corrections which can be reabsorbed into a redefinition of the PO.

\subsection[Prospects for the Higgs PO in ${Zh}$ at the HL-LHC]{Prospects for the Higgs PO in $\boldsymbol{Zh}$ at the HL-LHC}
\label{sec:Zh_prospects}

\begin{figure}
  \begin{center}
	\includegraphics[width=0.447\textwidth]{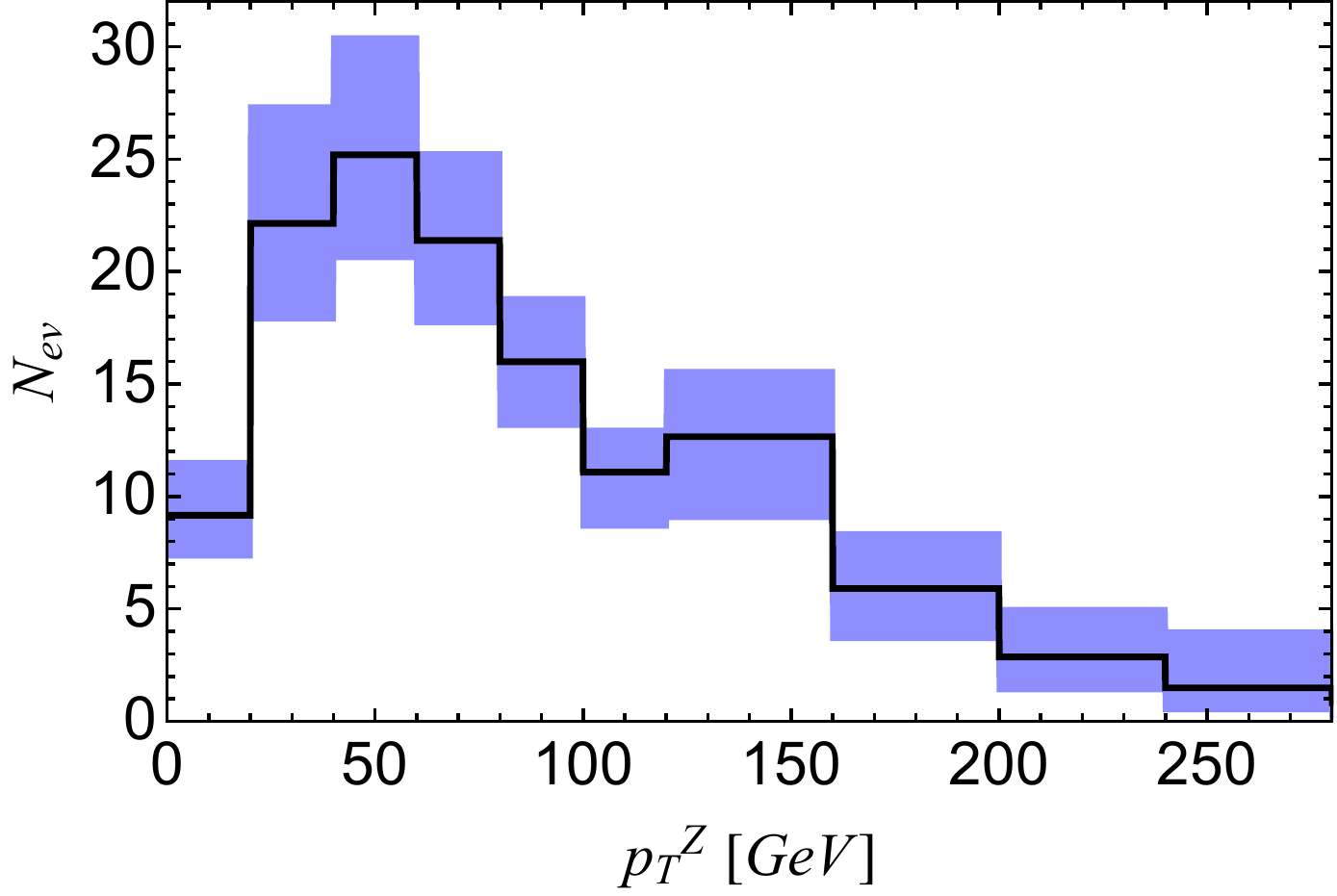} \quad
        \includegraphics[width=0.48\textwidth]{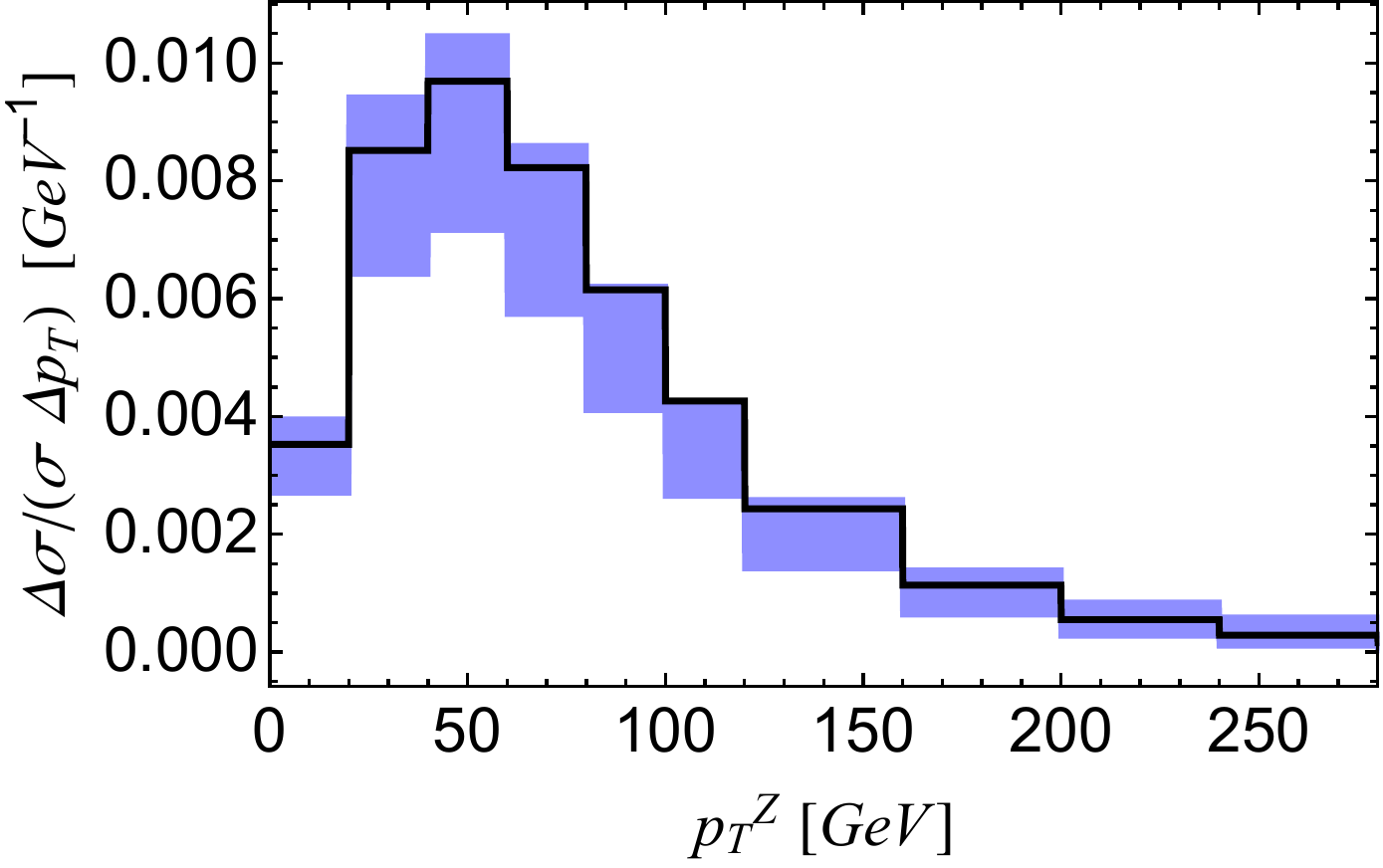}
  \end{center}
\caption{\small\label{fig:pTZ_2sigma_range} Allowed deviations in the $Z$ boson $\pT$ distribution by varying the PO within the $-2\log L / L_{\rm max} < 4$ ($2\sigma$) region. In the left plot we show the absolute number of events in each bin, while in the right one we show the normalized distribution with respect to the total number of events and the bin width.}
\end{figure}

In order to estimate the reach of the HL-LHC, at 13 TeV and $3000~ \text{fb}^{-1}$ of integrated luminosity, for measuring the Higgs PO in $Zh$ production, we consider the all-leptonic channel $Z \to 2\ell$, $h \to 2\ell2\nu$. 
The 8 TeV ATLAS search in this channel \cite{Aad:2015ona} estimated 0.43 signal events  with $20.3~\text{fb}^{-1}$ (Table X of \cite{Aad:2015ona}). By rescaling the production cross section and the luminosity up to the HL-LHC we estimate approximately $\sim 130$ signal events at the SM rate.
 Assuming a sample of this size we  perform a fit of the $\pT$ distribution of the $Z$ boson. In order to control the validity of the momentum expansion we  apply an upper cut of $\pT^{\rm max} = 280~ \GeV$, which corresponds approximately to $q^2 \approx 600~ \GeV$ (see Fig.~\ref{fig:pTZ_qSQ_SM}). We  bin the \pTZ distribution as $\{0-20-40-60-80-100-120-160-200-240-280\}~\GeV$. Using 
  the \UFO implementation of the PO within \Sherpa we generate $p p \to Zh$ events at $13~\TeV$ of c.o.m. energy. As in the VBF case, in each bin we have obtained the expression of the number of events as a quadratic function in the PO:
\be
	N_a^{\rm ev} =  \kappa^T X^a \kappa~, \quad \text{where} \quad
	\kappa = (\kappa_{ZZ}, \epsilon_{Zu_L}, \epsilon_{Zu_R}, \epsilon_{Zd_L}, \epsilon_{Zd_R})~,
\ee
where $a$  denotes again the label of each bin. We assume the number of events for each bin to follow a Poisson distribution
  and we  build the likelihood $L(\kappa)$ as a function of the five PO listed above. The best-fit point is defined by $L_{\textrm max}$ and we   determine $\Delta \chi^2 = -2\log L / L_{\rm max}$.
In Fig.~\ref{fig:HPO_HLLHC} we show the resulting $1\sigma$ ($2\sigma$) intervals for each PO with solid (dashed) blue lines, when all   other PO are profiled. The expected bounds obtained in the $Zh$ channel are comparable in strength with the ones obtained in the VBF channel.
In Fig.~\ref{fig:pTZ_2sigma_range} we  illustrate the $2\sigma$ allowed deviation of the $\pTZ$ distribution

A fit  based on a binning of the $Zh$ invariant mass spectrum provides very similar errors as those shown in Fig.~\ref{fig:HPO_HLLHC}.
 Again Gaussian errors obtained by expanding the likelihood as a quadratic function around the minimum overestimates the  
  errors compared to the ones shown in Fig.~\ref{fig:HPO_HLLHC}, although here not as badly as in the VBF case: 
\be
	Zh: \qquad \sigma^{\rm Gauss}_{\rm quad}(\kappa_{ZZ}, \epsilon_{Z u_L}, \epsilon_{Z u_R}, \epsilon_{Z d_L}, \epsilon_{Z d_R}) 
	= (0.085, 0.012, 0.014, 0.013, 0.019)\, .
	\label{eq:Zh_GaussFit_Quadr}
\ee
By multiplying the number of events in each bin by an overall rate modifier $\mu$, as done above for the VBF analysis, and profiling over this parameter, we find  $\kappa_{ZZ}$  being unconstrained but the $1\sigma$ errors 
 on the contact terms, in the Gaussian approximation, are exactly the same as the ones before. 
 This clearly implies that the bounds on the contact terms  arise from the shape information, and not from the rate.

\subsection[Prospects for the Higgs PO in $Wh$ at the HL-LHC]{Prospects for the Higgs PO in $\boldsymbol{Wh}$ at the HL-LHC}

In the case of $Wh$ production, in all the channels used for the Run-1 analysis, the signal manifests itself as a small excess over a 
large (dominating) background, see e.g.~Ref.~\cite{Aad:2015ona}. A detailed analysis for such processes should  be performed  evaluating carefully   the backgrounds, which is beyond the scope of this work. However, given the high luminosity we are looking at,  the golden channel $h\to 4\ell$, $W\to \ell \nu$ becomes an interesting viable possibility. It has been estimated by ATLAS that 67 signal SM events will be present with 3000~fb$^{-1}$ of integrated luminosity~\cite{Nisati:2015}.
We have thus decided to analyze  the prospects of this clean channel only, to constrain the $(\kappa_{WW}, \epsilon_{W u_L})$ PO,
with an analogous likelihood analysis as those performed for the $Zh$ and VBF  channels.

We have studied in particular the $\pTH$ distribution, as reference observable, applying the same binning and upper cut as in the $Zh$ analysis discussed above.
In Fig.~\ref{fig:HPO_HLLHC} we show the resulting $1\sigma$ ($2\sigma$) intervals for each PO with solid (dashed) green lines, when the other PO is profiled. In this case the Gaussian approximation works well and provides the following $1\sigma$ errors:
\be
	Wh: \qquad \sigma^{\rm Gauss}_{\rm quad}(\kappa_{WW}, \epsilon_{W u_L}) = (0.11, 0.0032)~.
\ee

Upon introducing a total rate modifier $\mu$, as done for the  previous channels, the bound on $\kappa_{WW}$ vanishes when $\mu$ is profiled.
However, the constraint on the contact-term PO $\epsilon_{W u_L}$ remains unchanged, implying that also in this case the bound 
 arises from the shape of the $\pT$ distribution.

\medskip

We conclude  the last two phenomenological sections stressing that we have performed simplified estimates of the HL-LHC sensitivity on the 
contact-terms PO by separately considering a limited set of collider signatures. It is reasonable to expect that, including all possible signatures and performing a global fit, the sensitivity  can significantly  improve. However, such a global analysis should also consider the effect of backgrounds, neglected in this study.

\section{Validity of the momentum expansion}
\label{sec:mom_exp}
\label{sect:expansion}

The most important check to estimate the validity of the momentum expansion is represented by 
the consistency condition (\ref{eq:consistency}), where $q^2_{\rm{max}}$ is controlled  
by $(\pTj)^{\rm max}$ in VBF and $m_{Vh}$ in VH (or, less efficiently, by $\pTZ$ and $\pTH$ in VH). 
Besides checking this condition 
a further  check to assess the validity of the momentum expansion is obtained comparing the fit performed including the full quadratic dependence of $N^{\rm{ev}}_a$ on the PO, with a fit in which the $N^{\rm{ev}}_a$ are linearized in $\delta \kappa_X \equiv \kappa_X-\kappa_X^{\rm{SM}}$ and $\epsilon_X$. The idea behind this procedure is that the quadratic corrections to physical observable in $\delta \kappa_X$ and $\epsilon_X$ are  formally of the same order as the interference of the first neglected term in Eq.~\eqref{eq:FLGL} with the leading SM contribution.

If the two fits (linear vs.~quadratic) 
provide similar results, one can safely conclude that  the terms neglected in the PO decomposition are indeed subleading.
In principle, if the two fits yield significantly different results, the difference might be used to estimate the uncertainty due to the neglected 
higher-order terms in the momentum expansion. In practice, as will be illustrated below, this estimate turns out to be rather 
pessimistic and often an overestimate of the uncertainty on the PO. 

To access the feasibility of this check, we perform a linear fit for the VBF Higgs production closely following the procedure described in Sec.~\ref{sec:VBF_prospect}.
The results obtained in  the Gaussian approximation are:
\be\begin{split}
	{\rm VBF:} \qquad \sigma^{\rm Gauss}_{\rm linear}(\delta\kappa_{ZZ}, \delta \kappa_{WW}, \epsilon_{Z u_L}, \epsilon_{Z u_R}, \epsilon_{Z d_L}, \epsilon_{Z d_R}, \epsilon_{W u_L}) =& \\
	= (1.7, 0.42, 0.30, 0.57, 0.32, 1.0, 0.038)~. &
	\label{eq:VBF_GaussFit_Lin}
\end{split}
\ee
Comparing those results with Eq.~\eqref{eq:VBF_GaussFit_Quadr}, we conclude that the bounds on the contact terms in the linearized case are
significantly weaker (typically one order of magnitude less stringent) than those obtained in the quadratic fit. Similar results are obtained for the $Zh$ analysis, while only in the $Wh$ case the two fits give comparable results:
\be\begin{split}
	Zh: \qquad & \sigma^{\rm Gauss}_{\rm linear}(\delta\kappa_{ZZ}, \epsilon_{Z u_L}, \epsilon_{Z u_R}, \epsilon_{Z d_L}, \epsilon_{Z d_R}) 
	= (0.2, 0.14, 0.32, 0.11, 0.35)~,  \\
	Wh: \qquad &\sigma^{\rm Gauss}_{\rm linear}(\delta\kappa_{WW}, \epsilon_{W u_L}) = (0.11, 0.0033)~.
	\label{eq:Vh_GaussFit_Lin}
\end{split}
\ee 
Given the events we have simulated are obtained using SM-like distributions, we cannot attribute this large 
difference to a possible breakdown of the momentum expansion in the underlying distribution. 
We dedicate the rest of this section to investigate in more detail the origin of the mismatch and how to address it. 

The most likely explanation for the large difference between linear and quadratic fits  reported above is  the fact that in the linear fit only a few linear combinations of the PO enter the observables, thus reducing the number of independent constraints one can get. This fact, coupled to the large number 
of free parameters in VBF and $Zh$, could explain the loose constraints obtained in the linear fit. 
If this was true, we should find that in simple models with less parameters the linear and quadratic fit should agree. 

To check if the constraints obtained on the contact terms can, in fact, be used to bound explicit new physics scenarios, we employ 
a simple toy model. To this end, we extent the SM with a new neutral vector boson, $Z'$, coupled to specific fermion currents (to be defined below) and to the Higgs, such 
that it contributes to VBF and VH (or better $Zh$) production.
Since the goal of this section is to examine the validity of the momentum expansion with an explicit new physics example, we ignore all other phenomenological constraints on such a model (for example, electroweak precision tests, direct searches, etc).\footnote{~For recent studies about the validity of the momentum expansion in VBF and $Zh$ using similar toy models see Ref.~\cite{Brehmer:2015rna,Biekoetter:2014jwa}).}

One the one hand, we compute the bounds on the mass and couplings of this new state from the analysis of the double differential $\pT$ distribution in VBF Higgs production (and the $\pT^Z$ distribution in $Zh$). On the other hand, we integrate out the heavy $Z'$ and match to the Higgs PO framework. Finally, we compare the bounds in the full model with the ones obtained from the Higgs PO fit.

To be more specific, we consider a $Z'$ which contributes to the form factor $F_L^{f f'}$ of $\langle J_f^\mu(q_1) J_{f'}^\nu(q_2) h \rangle$ as
\beq
	F_L(q_1^2, q_2^2)^{f f'} = F_{L, {\rm SM}}^{f f'}(q_1^2, q_2^2) -  \frac{v}{m_Z} g_H \left[ \frac{ g_{Z'}^{f} g_Z^{f'} }{P_{Z'}(q_1^2) P_{Z}(q_2^2)} + \frac{ g_{Z}^{f} g_{Z'}^{f'} }{P_{Z}(q_1^2) P_{Z'}(q_2^2)} \right]~,
	\label{eq:ZpFormFactor}
\eeq
Such a contribution could arise, for example, from the following interaction terms,
\beq
	\Lag \supset - 2 g_H m_Z Z^\mu Z'_\mu h + \sum_{f=f_L,f_R} g_{Z'}^{f} \bar{f} \gamma^\mu f Z'_\mu~,
	\label{eq:Zprime_int}
\eeq
 where all the fields are canonically normalized and in the mass basis.  
 Using \FeynRules ~\cite{Christensen:2008py} (package version 1.6.16) we obtain an  \UFO~\cite{Degrande:2011ua} representation of this $Z'$-model and perform exactly the same analysis previously applied to the PO for VBF and \VH~production. This allows us to derive bounds  on the combination of couplings $g_f \equiv g_H g_{Z'}^{f}$ for a set of benchmark $Z^\prime$ masses, $M_{Z'}$. In this simple model
 the $Z'$ only decays to a pair of fermions as well in $Z+h$. The corresponding partial decay widths, assuming the $Z'$ is much heavier than the daughter particles, are
\be
\Gamma(Z' \to \bar f f) = \frac{N_c ~M_{Z'}}{24 \pi} |g_{Z'}^{f}|^2~, \qquad  \Gamma(Z' \to Z h) =\frac{M_{Z'}}{48 \pi} g_H^2~, 
\label{eq:Zprime_width}
\ee
where $N_c$ is the number of colors. In order to simplify the analysis, we assume that the $Z'$ is a narrow resonance 
($\Gamma_{Z'} \ll M_{Z'}$). This allows to interpret bounds from the VBF and VH analyses in terms of the $g_f$ parameters. Using the above relations, we have checked that this condition is satisfied for the benchmark scenarios we consider in the following.  
Expanding the form factor from Eq.~\eqref{eq:ZpFormFactor} for $q_{1}^2 \ll M_{Z'}^2$ and  $\Gamma_{Z'} \ll M_{Z'}$  and keeping only the leading deviation from the SM, we find:
\beq
	\epsilon_{Z f} =  g_H g_{Z'}^{f} \frac{v m_Z}{M_{Z'}^2} = g_f \frac{v m_Z}{M_{Z'}^2} ~.
	\label{eq:ZpHPOmatch}
\eeq

%

\subsection[Effect of the $Z'$ in VBF]{Effect of the $\boldsymbol{Z'}$ in VBF}

\begin{figure}[t]
  \begin{center}
    \includegraphics[width=0.45\textwidth]{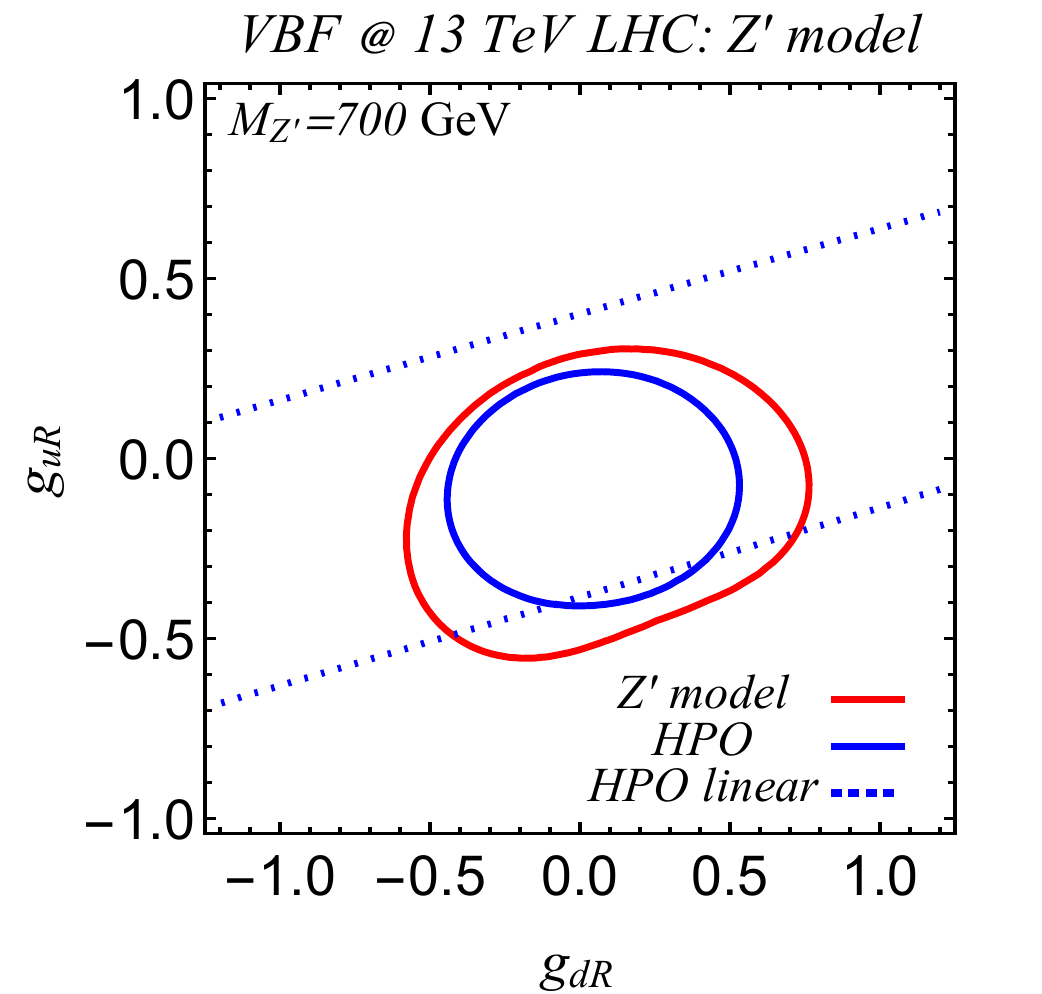} \quad 
    \includegraphics[width=0.428\textwidth]{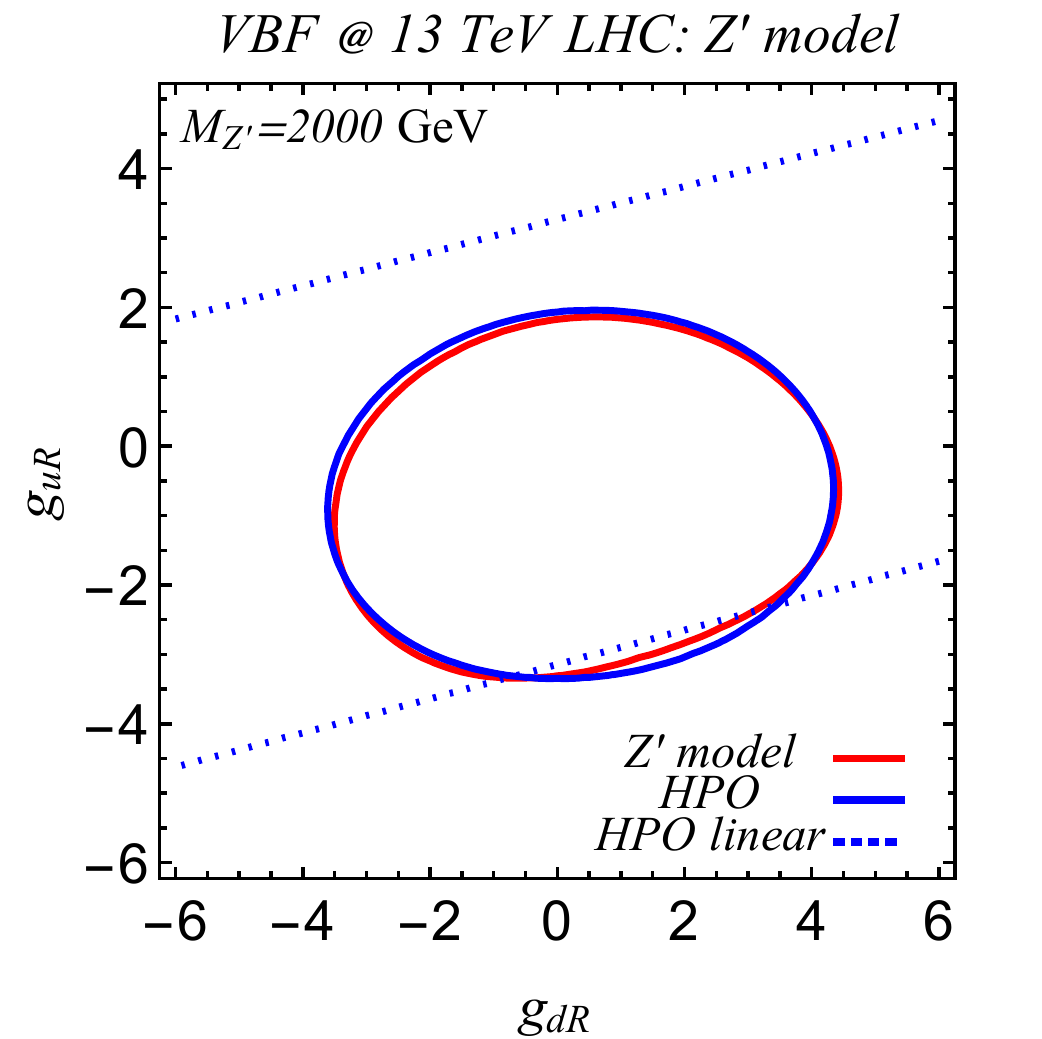}
  \end{center}
\caption{\small\label{fig:Zpbound_2d} We show the expected 95 \% CL bound in the plane $(g_{d_R}, g_{u_R}) \equiv g_H (g_{Z'}^{d_R}, g_{Z'}^{u_R})$ for $M_{Z'} = $ 700  and 2000 GeV on the left and right plots, respectively. All the bounds are obtained analysing 2000 VBF Higgs production events as discussed in Sec.~\ref{sec:VBF_prospect}. The solid red line represents the bound obtained in the $Z'$ model, while the solid blue (dotted blue) are the bounds obtained in the Higgs PO fit with quadratic (linear) dependence on the PO. }
\end{figure}
 
We consider the case where the $Z'$ couples to both the down and up right-handed quarks, with two independent couplings, $g_{Z'}^{d_R}$ and $g_{Z'}^{u_R}$. In addition, we fix the $Z'$ mass to two benchmarks values: (a) $700$~GeV and (b) $2000$ GeV.
The main results of the analysis are shown in Fig.~\ref{fig:Zpbound_2d}. 

On the one hand, we perform a fit to the Higgs PO $\epsilon_{Z u_R}$ and $\epsilon_{Z d_R}$,  while fixing all other PO to zero,  and translate this bound on the relevant parameter space of the $Z'$ model, namely the $\{g_{d_R}, g_{u_R}\}$ plane. We report the results of the fit obtained with full quadratic dependence on the PO, as well as the results in which $N_{\rm{ev}}$ is linearized in $\delta \kappa_X$ and $\epsilon_X$. In both cases, $95\%$ CL bounds are obtained by requiring $-2\log L / L_{max} \le 5.99$. 
On the other hand,   using exactly the same binning and statistical treatment, we directly fit the $Z'$ model parameters.

Comparing the two methods we conclude: (i) for both masses the quadratic PO fit provides a reasonable approximation of the model fit, while the 
linear fit largely overestimates the errors; (ii) the PO fit performs better for $M_{Z'} = 2000$~GeV than for $M_{Z'} =700$~GeV, as expected from the momentum expansion validity arguments (we recall that we set the cut $\pTj < 600$~GeV); however, also for  $M_{Z'} = 700$~GeV the quadratic fit does provide a fair approximation to the model fit. In particular, in this case we see that the bound from the PO fit is stronger than in the model, which can be understood by the fact that in VBF the $Z'$ is exchanged in the $t$-channel, and therefore its main effect is to reduce the amplitude for high values of $q^2$.

\subsection[Effect of the ${Z'}$ in $Zh$]{Effect of the $\boldsymbol{Z'}$ in $Zh$}

\begin{figure}[t]
  \begin{center}
    \includegraphics[width=0.455\textwidth]{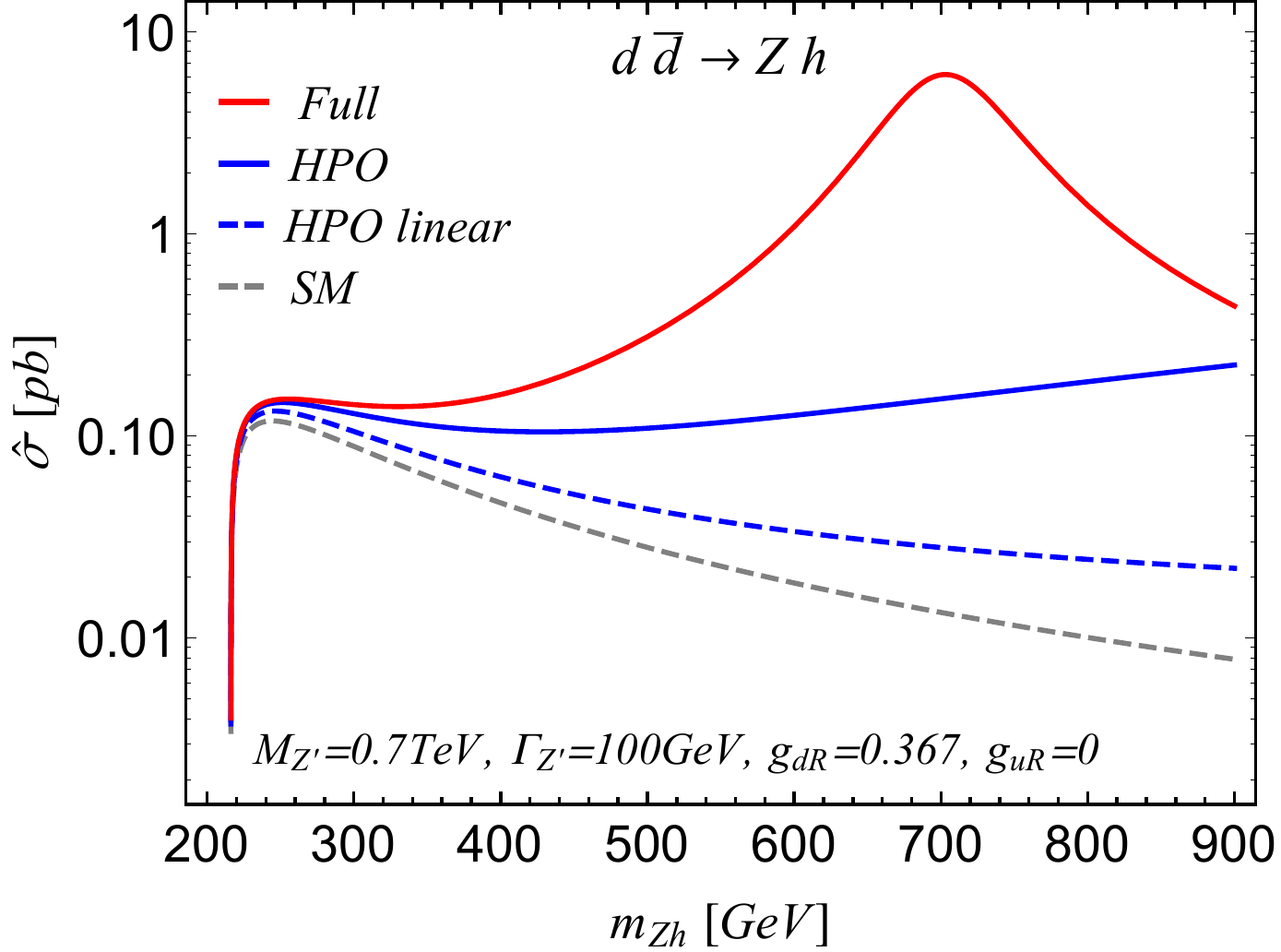} \quad 
    \includegraphics[width=0.47\textwidth]{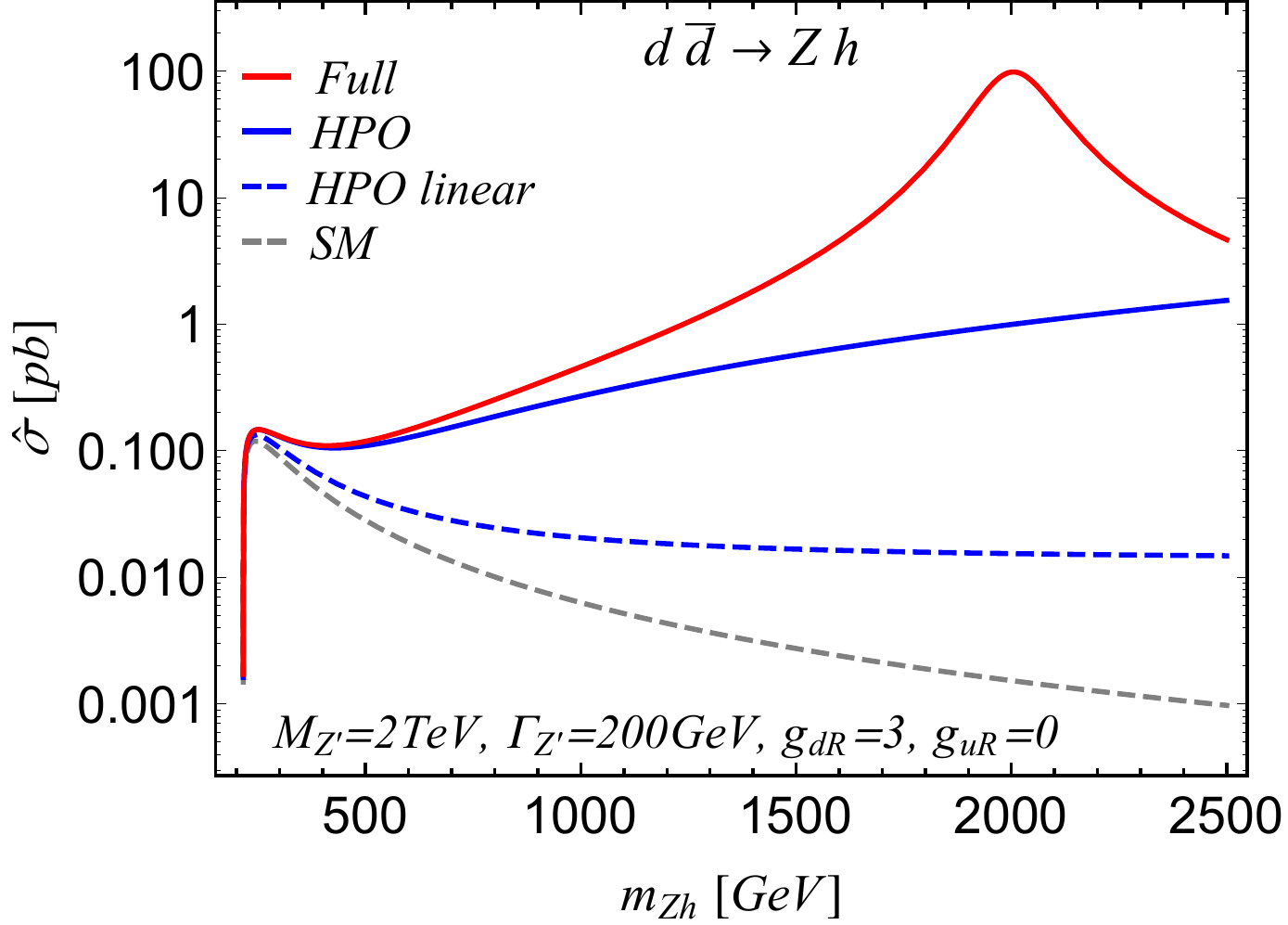}
  \end{center}
\caption{\small\label{fig:Zp_Zh_partonic} Partonic cross section $d \bar{d} \to Z h$ as a function of the invariant mass $m_{Zh}$ in the SM (dashed gray line) and with a $Z^\prime$ coupled to right-handed down quarks only. With red lines we show the cross section computed in the full model while the blue ones represent the cross section using the PO decomposition --with matching conditions in
Eq.~\eqref{eq:ZpHPOmatch}-- using the full dependence (solid line) or only the linear one (dashed line). In the left plot we consider the benchmark light
 $Z^\prime$ scenario:  $M_{Z'} = 700~\GeV$, $\Gamma_{Z'} = 100~ \GeV$ and $g_{d_R} = 0.367$. In the right plot we consider the heavy $Z'$ scenario: $M_{Z'} = 2000~\GeV$, $\Gamma_{Z'} = 200~ \GeV$ and $g_{d_R} = 3$. Both benchmarks give rise to the same contact term: $\epsilon_{Z d_R} \simeq 1.68 \times 10^{-2}$.}
\end{figure}

In order to assess the validity of the momentum expansion in associated production, it is convenient to look first at the underlying partonic cross section. In Fig.~\ref{fig:Zp_Zh_partonic} we show the partonic cross section $d \bar{d} \to Z h$, 
as a function of the $Zh$ invariant mass, for the two benchmark points of $Z'$ model introduced above. 

Both benchmark points have been chosen such that they generate the same contact term when the $Z'$ is integrated out, $\epsilon_{Z d_R} = 1.68 \times 10^{-2}$, which is within the $2\sigma$ bound of our PO fit.
The width of the $Z'$ has been fixed to $100~\GeV$ and $200~\GeV$ for the light and heavy scenario, respectively. Using Eq.~\eqref{eq:Zprime_width} and assuming no other decay mode is present, this corresponds to $g_H \simeq 0.097 ~ (3.0)$ in the light (heavy) scenario. We have checked that our conclusions do no change by varying the total width, as long as the condition $\Gamma_{Z'} \ll M_{Z'}$ is satisfied.

\begin{figure}[t]
  \begin{center}
    \includegraphics[width=0.45\textwidth]{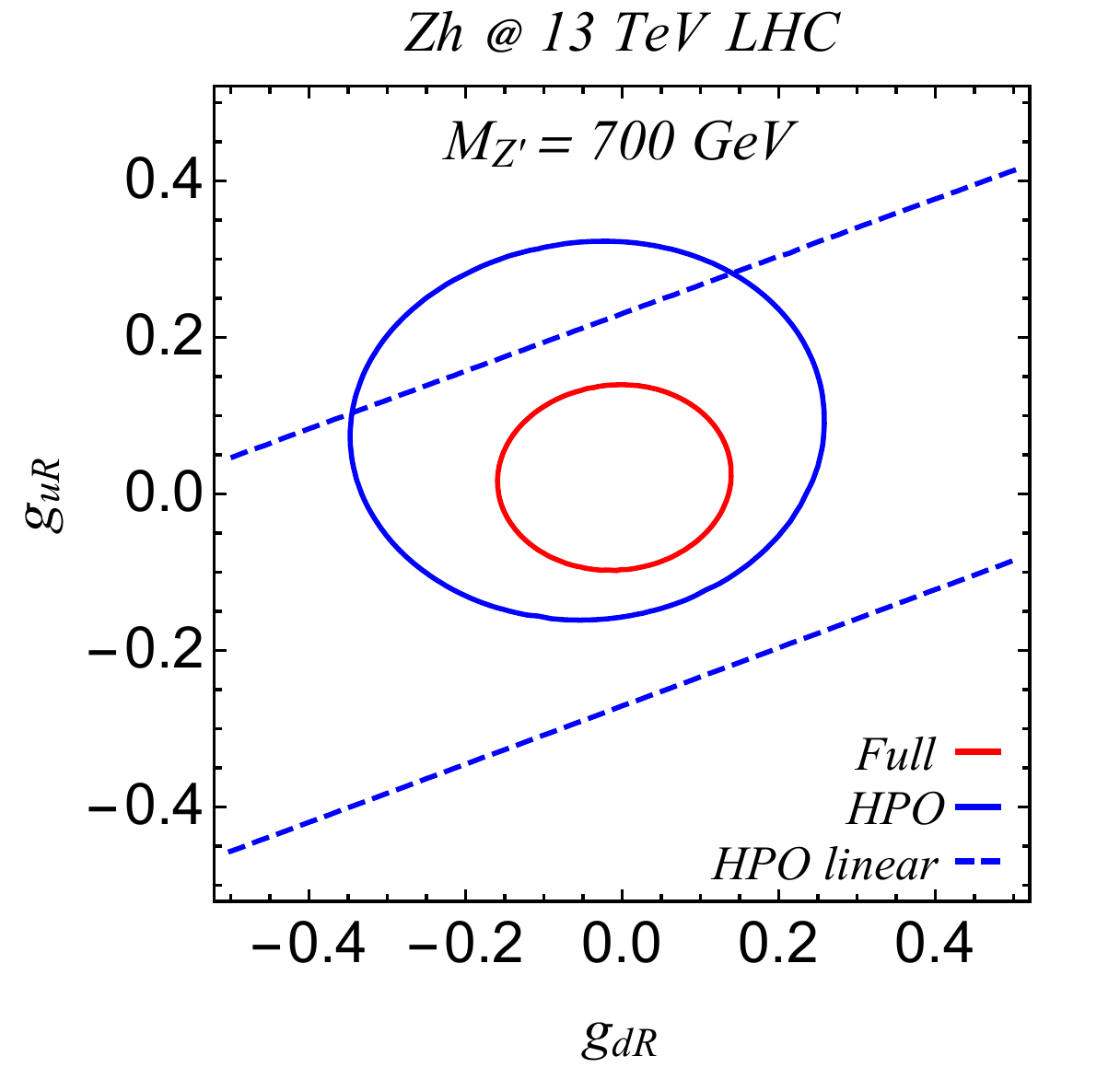} \quad 
    \includegraphics[width=0.43\textwidth]{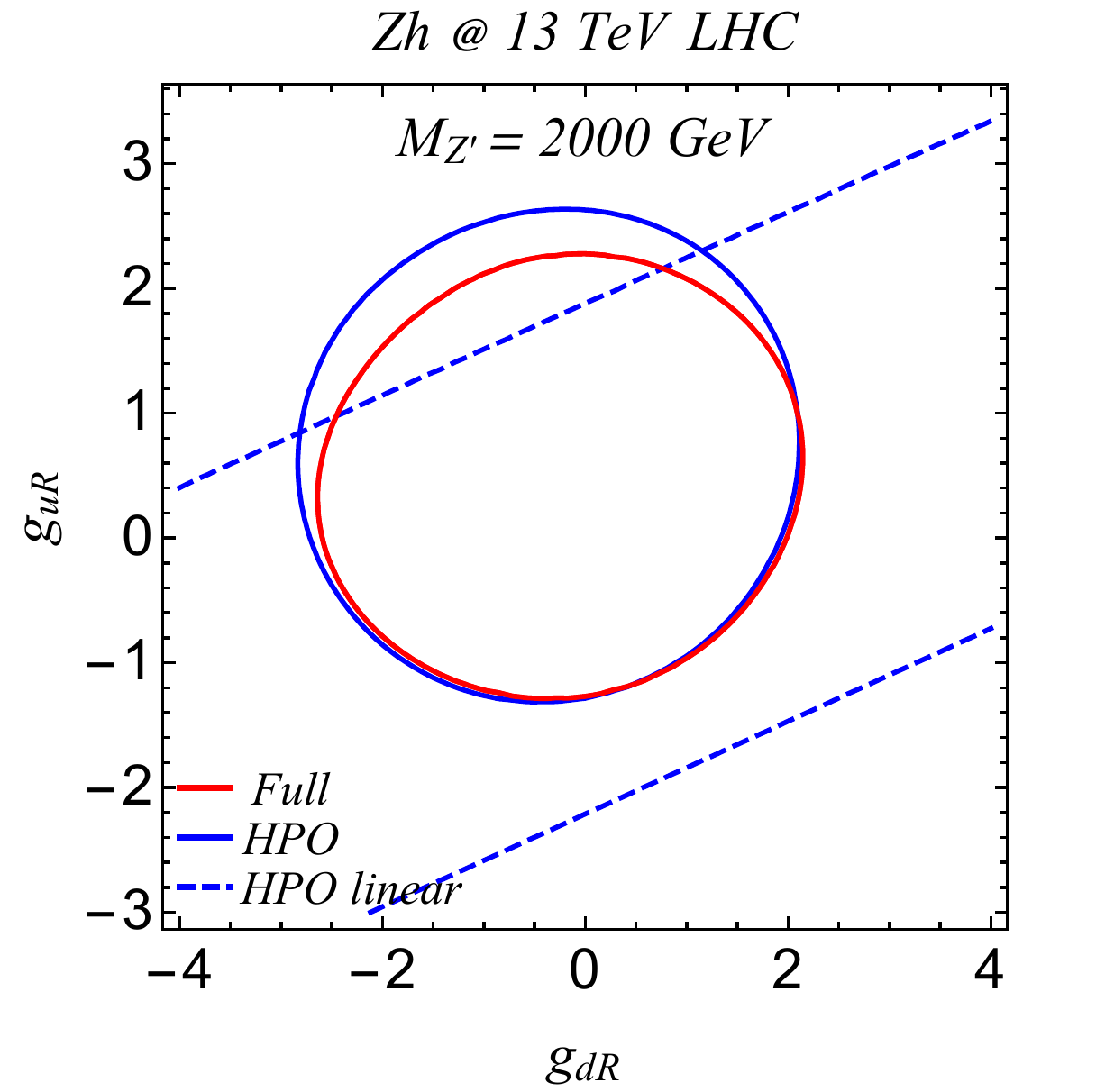}
  \end{center}
\caption{\small\label{fig:Zpbound_2d_Zh}  Expected 95 \% CL bound in the plane $(g_{d_R}, g_{u_R}) \equiv g_H (g_{Z'}^{d_R}, g_{Z'}^{u_R})$ for $M_{Z'} = $ 700  and 2000 GeV on the left and right plots, respectively. All the bounds are obtained analysing 130 $Zh$ Higgs production events as discussed in Sec.~\ref{sec:Zh_prospects}. The solid red line represents the bound obtained in the full model, while the solid blue (dashed blue) are the bounds obtained via the matching in eq.~\eqref{eq:ZpHPOmatch} from the Higgs PO fit with quadratic (linear) dependence on the PO.}
\end{figure}

As expected, in the light scenario the cross section in the full model strongly deviates from the PO one well before the $600~\GeV$ cutoff imposed in the fit, implying that our PO fit is not reliable in this case.
On the other hand, the scenario with a heavy and strongly coupled $Z'$ shows a very good agreement with the full PO analysis up to $\sim 1~\TeV$, i.e. well above the UV cutoff of our analysis, implying that the analysis can be safely applied to such scenarios, and that it could be even improved by setting a slighly higher cutoff.
In both cases, from Fig.~\ref{fig:Zp_Zh_partonic} is clear that the linearized dependence on the PO is not sufficient to describe the cross section, even for energies much smaller than the $Z'$ mass.

From this analysis we can anticipate the results of a comparison of various fits of $Zh$ data, i.e.~full model fit vs.~PO fits using quadratic and linear dependence, as already done in the VBF case. In Fig.~\ref{fig:Zpbound_2d_Zh} we show the results of such fits. We stress that in all cases the analysis was exactly the same: we have analyzed  the $\pT^Z$ distribution up to $280~\GeV$, employing always the same binning 
(as discussed in Sec.~\ref{sec:Zh_prospects}). The solid red line represents the 95 \% CL bound in the full model while the solid (dashed) blue line shows the bound obtained from the PO fit with quadratic (linear) dependence. 

The distributions in Fig.~\ref{fig:Zp_Zh_partonic} allow a straightforward interpretation of these results. In the heavy-$Z^\prime$ case, the full quadratic expansion in the Higgs PO describes very well the $m_{Zh}$ distribution before the cutoff of $600~\GeV$, while keeping only the linear dependence underestimates the new physics contribution. It is thus expected that in this case the bound will be much worse.
In the light-$Z^\prime$  case,  both expansions with Higgs PO underestimate the cross section, thus providing a worse bound than in the full model. Still, the quadratic dependence does a significantly better job in approximating the complete model than the linear one, as in the VBF case. 

\medskip

From this illustrative toy-model example  we can draw the following general conclusion with respect to the validity of the PO expansion: 
for underlying models that respect the momentum expansion, hence for models where the PO extracted from 
data satisfy, a posteriori,  the consistency condition (\ref{eq:consistency}), the 
quadratic fit provides more reliable and thus more useful constraint on the PO. In such models the difference between quadratic and linear fit represents a large overestimate of the errors.

However, the situation is more involved  for models with low-scale new physics. The latter should manifest by anomalously 
large values of the PO, or sizable differences in the fits performed with different upper $\pT$ cuts. In such cases   
the quadratic fit is likely to provide a useful constraint, especially for the class of models with a strong correlation between 
linear and quadratic terms in the momentum expansion (as the simple $Z^\prime$ model discussed above). 
Still, for low-scale new-physics we cannot   exclude more complicated  scenarios where new model parameters appearing 
at higher order in the momentum expansion wash-out an apparent small error on the PO from the quadratic fit.
In such cases only the the results of the linear fit (with a properly low $\pT$ cut) would provide an unbiased constraint on the model. 

In view of these arguments, we encourage the experimental collaborations to report the results of both linear and quadratic fits,
as well as to perform such fits using different  $\pT$ cuts.

\section{Conclusions }
\label{sect:Conc}

Higgs physics is entering the era of precision measurements: future
high-statistics data will allow us not only to determine the overall signal
strengths of production and decay processes relative to the SM, but also to
perform detailed kinematical studies. In this perspective, an accurate and
sufficiently general parameterization of possible NP effects in such
distributions is needed. In this paper we have shown how this goal can be
achieved in the case of VBF and VH production, generalizing the concept of Higgs
PO already introduced in Higgs decays.
 
As summarized in Table~\ref{tab:POsumm}, the number of additional PO appearing
in all VBF and VH production amplitudes is manageable.
In particular, assuming CP invariance, flavor and custodial symmetry,
only 4 new PO should be added to the set of 7 PO appearing in $h\to 4\ell, 2\ell
2\nu, 2 \ell \gamma, 2\gamma$ in the same symmetry
limit~\cite{Gonzalez-Alonso:2014eva}. This opens the possibility of precise global determinations of the PO from combined analyses
of production and decay modes, already starting from the next LHC runs.
 
As extensively illustrated in Sections~\ref{sect:VBF} and \ref{sect:VH}, the key
aspects of VBF and VH is the possibility of exploring sizable momentum transfers
in the Green functions of Eq.~(\ref{eq:corr_func}). On the one hand, this
maximizes the sensitivity of such processes to PO that are hardly accessible in
Higgs decays. On the other hand, it allows us to test the momentum expansion
that is intrinsic in the PO decomposition as well as in any EFT approach to
physics beyond the SM.
Key ingredients to reach both of these goals are precise differential measurements of
 ${\rm d}^2\sigma/{\rm d}\pTjone {\rm
d}\pTjtwo$ in VBF and ${\rm d}\sigma/{\rm d} m_{Vh}$ in VH
(or appropriate proxies such as
$\pTH$ and $\pTZ$). We thus encourage
the experimental collaborations to directly report such differential
distributions, especially in the kinematical regions corresponding to high
momentum transfer.

As far as the PO fits in VBF an VH are concerned, we suggest to perform them
setting a maximal cut on $\pTj$ and $m_{Vh}$, to ensure (and verify a posteriori) the validity of the
momentum expansion. As illustrated by matching the PO framework to simplified
dynamical NP models, it is also important to report the results of fits using
both linearized and quadratic expressions for the cross-sections in terms of PO.
According to our preliminary estimates, the production PO could be measured at
the percent level at the HL-LHC (in the case of maximal flavor symmetry, without
the need of imposing custodial symmetry). This level would be sufficient to
constrain (or find evidences) of a wide class of explicit NP models and, among
other things, to perform non-trivial tests of the relations between electroweak
observables and Higgs PO expected in the SMEFT.

\subsection*{Acknowledgements}

We would like to thank S. H\"oche and S. Kuttimalai for help with the UFO interface of Sherpa. Also we would like thank 
M. Duehrssen-Debling, S. Pozzorini, and A. Tinoco Mendes for useful discussions.
This research was supported in
part by the Swiss National Science Foundation (SNF) under contract 200021-159720.


\appendix


\section{Details on the statistical analysis}
\label{app:stat}

In this Appendix we provide details on the statistical analysis used to derive the projected sensitivity on the PO. The first step of such an analysis is to compute, by means of Monte Carlo simulation, the signal yield in each bin as a quadratic polynomial in the PO:
\begin{equation}
	N^{\rm ev}_a = \kappa^T X^a \kappa~,
\end{equation}
where $a$ labels a given bin. For example, the total cross section for VBF Higgs production at 13 TeV, applying the cuts defined in Section~\ref{sec:VBF_prospect}, is given by
\be
	\frac{\sigma_{\rm VBF}^{\rm PO}}{\sigma_{\rm VBF}^{\rm SM}} = \kappa^T \left(
\begin{array}{ccccccc}
 0.32 & 0.02 & -6.2 & 3.29 & 5.68 & -0.72 & 0. \\
 0. & 1.06 & 0. & 0. & 0. & 0. & -25.3 \\
 0. & 0. & 122 & -15.0 & -27.1 & 4.92 & -1.48 \\
 0. & 0. & 0. & 108 & 12.2 & -2.1 & 0. \\
 0. & 0. & 0. & 0. & 72 & -3.72 & 1.01 \\
 0. & 0. & 0. & 0. & 0. & 61.6 & 0. \\
 0. & 0. & 0. & 0. & 0. & 0. & 325 \\
\end{array}
\right) \kappa~.
\ee
After deriving similar expression for each bin, we perform a profile likelihood fit to a binned histogram distribution. The likelihood function takes the form
\begin{equation}
- 2 \log L = \sum_{a\in{\rm bins}} \left( -2 \log \left( \frac{\exp[-\kappa^T \tilde{X}^a \kappa] ~ (\kappa^T \tilde{X}^a \kappa)^{\Delta N^{\rm exp}_a}}{\Delta N^{\rm exp}_a !}\right)+\sum_{i j} \left(\frac{\tilde{X}^a_{i j}-X^a_{i j}}{\Delta X^a_{i j}} \right)^2 \right)~,
\end{equation}
where $\Delta N^{\rm exp}_a$ denotes the number of projected-observed events in a bin with label $a$ (which we take to be SM-like),  $X^a_{i j}$ are the coefficients of the $X^a$ matrix as obtained from our simulation and $\Delta X^a_{i j}$ are the uncertainties associated with these coefficients. These uncertainties are determined from a Poisson distribution in the number of events in each bin and from a normal distributions in the nuisance parameters $\tilde{X}^a_{i j}$. We first minimize the above function with respect to the PO ($\kappa$) and nuisance parameters ($\tilde{X}^a_{i j}$) and then expand the function around the best fit point up to second order
\be
- 2 \log L - (- 2 \log L)_{\rm min} \approx \Delta \chi^2 = (\kappa-\kappa_{\rm min})^T V^{-1} (\kappa-\kappa_{\rm min})+\dots~,
\ee
where dots represent terms that involve the nuisance parameters as well. Here, $V_{i j} = \sigma_i \rho_{i j} \sigma_j$ where $\sigma_i$ and $\rho_{i j}$ are the uncertainties and correlation coefficients, respectively.
We refer to this method as the Gaussian approximation, and use it to study the impact due to Monte Carlo uncertainties, as well as due to missing higher order corrections on the fit results. On the other hand, the results shown in Fig.~\ref{fig:HPO_HLLHC} are obtained by setting the error on the nuisance parameters to zero, and for each PO, profiling over all the others.

\bibliographystyle{JHEP}
\bibliography{biblio}
\end{document}